\documentclass[a4paper,fleqn]{cas-dc}

\usepackage{float}
\usepackage[numbers,sort&compress]{natbib}
\usepackage{siunitx}
\usepackage{miller}
\usepackage{graphicx}
\usepackage{caption}
\usepackage{subcaption}

\DeclareSIUnit\bar{bar}

\def\tsc#1{\csdef{#1}{\textsc{\lowercase{#1}}\xspace}}
\tsc{WGM}
\tsc{QE}
\tsc{EP}
\tsc{PMS}
\tsc{BEC}
\tsc{DE}

\begin{document}
\let\WriteBookmarks\relax
\def\floatpagepagefraction{1}
\def\textpagefraction{.001}

\emergencystretch 3em

\shorttitle{Capturing thin-film microstructure contributions during ultrafast laser-metal interactions}
\shortauthors{Ganesan \& Sandfeld}

\title[mode = title]{Capturing thin-film microstructure contributions during ultrafast laser-metal interactions using atomistic simulations}

\author[1]{Hariprasath Ganesan}[type=editor,
                        auid=000,bioid=1,
                        orcid=0000-0002-2152-5928]
\cormark[1]
\ead{h.ganesan@fz-juelich.de}
\ead[url]{www.fzj.de}

\credit{Conceptualization of this study, Methodology, Software, Writing - Original draft preparation}

\address[1]{Institute for Advanced Simulations – Materials Data Science and Informatics (IAS‑9), Forschungszentrum Juelich, Juelich, Germany}

\author[1,2]{Stefan Sandfeld}[type=editor,
                        auid=000,bioid=2]
\credit{Funding Acquisition, Project Administration, and Writing}

\address[2]{RWTH Aachen University, Faculty of Georesources and Materials Engineering,
Chair of Materials Data Science and Materials Informatics, Aachen, Germany}

\cortext[cor1]{Corresponding author}

\begin{abstract}
Progress in the emerging fields of atomic and close-to-atomic scale manufacturing is underpinned by enhanced precision and optimization of laser-controlled nanostructuring.
Understanding thin films’ crystallographic orientations and microstructure effects becomes crucial for optimizing the laser-metallic thin film interactions; however, these effects remain largely unexplored at the atomic scale. 
Using a hybrid two-temperature model and molecular dynamics, we simulated ultrafast laser-metal interactions for gold thin films with varying crystallographic orientations and microstructure configurations.
Microstructure features, namely grain size, grain topology, and local crystallographic orientation, controlled the rate and extent of lattice disorder evolution and phase transformation, particularly at lower applied fluences. Our simulations provided comprehensive insights encompassing both the nanomechanical and thermodynamic aspects of ultrafast laser-metal interactions at atomic resolution.
Microstructure-aware/informed thin film fabrication and targeted defect engineering could improve the precision of nanoscale laser processing and potentially emerge as an energy-efficient optimization strategy.
\end{abstract}

\begin{keywords}
femtosecond laser \sep Au thin-film  \sep laser-metal interaction \sep molecular dynamics \sep microstructure
\end{keywords}

\maketitle
\section{Introduction}

Ultrafast laser structuring of materials and nano-machining of surfaces emerge as high-precision yet energy-efficient processing techniques, accelerating progress in atomic and close-to-atomic scale manufacturing \cite{li2022shaped,wang2023laser}, finding applications in various materials from dielectrics, semiconductors, and metals \cite{shugaev2016fundamentals}.
Progress in metal machining precision (sub-nm range) and quality \cite{lenzner2000photoablation} demands a profound understanding of the ultrafast (femtosecond) laser-metal interactions \cite{xiong2017effect}.
Metallic thin films are nanometer ($\si{\nano\metre}$) to micrometer-scale ($\si{\micro\metre}$) layers of high-purity materials, finding a range of technological applications exploiting their electronic, magnetic, optical, and mechanical properties \cite{richard1996mechanical}.
The consideration of crystallographic orientations and microstructure defects in thin films becomes inevitable due to two scenarios: 
1.) Thin films are fabricated using physical/chemical deposition techniques, and they remain susceptible to preferred crystallographic orientations and microstructure defects owing to the substrate choice and built-in stresses during thin film production. 2.) These thin films undergo microstructure modification during multi-pulse laser irradiation due to superheating and undercooling.
To this end, thin films' crystallographic orientations and microstructures/defects determine their effective target applications (i.e., mechanical, electrical, and optical properties) and processing precision \cite{lenzner2000photoablation} due to laser-metal interaction.
Consequently, finding the optimal process conditions and parameters of laser-controlled nanostructuring requires several time-consuming and energy-intensive trial-and-error experiments.

Laser-metal interaction remains the core phenomenon of these process techniques, which confines high energy in small regions of the irradiated material at an ultrafast time. 
\cite{shugaev2016fundamentals} categorize such interactions into three stages: 1.) laser excitation and non-thermal process, 2.) laser-induced structural and phase transformation, and 3.) laser ablation encompassing different activated mechanisms.
Deposited high energy within ultrafast time drives the system into an electronic, thermodynamic, and mechanical non-equilibrium state characterized by different electron and ion temperatures.
Subsequently, after the laser pulse duration, the deposited energy transfers from the electronic to the lattice subsystem, resulting in extreme heating ($10^{14}$\thinspace$\si{\kelvin/\second}$) of the target material \cite{shugaev2016fundamentals}. 
Such femtosecond laser pulse induces nanoscale phenomena in absorbing material without any substantial/recognizable material damage \cite{zhakhovskii2009molecular}.
Several pioneering works reported evidence and insights related to such nanoscale phenomena \cite{sokolowski1998transient,siwick2003atomic,sokolowski2003femtosecond,widmann2004single,rafaja2004interference}.
However, the experimental methods remain challenging in observing the complete dynamics of laser-induced nanoscale processes directly at atomic resolution.

Investigating ultrafast laser-metal interactions encompasses events that are extremely fast in time (fs-ps) and at smaller length ($\si{\nano\metre}$-$\si{\micro\metre}$) scale, thus making atomistic simulation method like \textit{classical} molecular dynamics (MD) an obvious choice in accessing the \textit{in silico} spatiotemporal information at atomic resolution. 
MD treats the material system as a set of discrete particles/ atoms/ lattice ions and simulates their collective behavior under initial and boundary conditions \cite{ganesan2018parallelization,ganesan2021parallel}.
MD provides the time trajectory of such materials systems by numerically integrating Newton’s equation of motion and captures the complex interplay of atomic system’s thermal, mechanical, and chemical contributions \cite{ganesan2021quantifying,ganesan2021understanding,ganesan2024modeling}. 
However, high energy deposition in metals during ultrafast laser-metal interaction absorbed by electrons results in a thermal non-equilibrium state between hot electrons and cold lattice subsystems.
Thus, we require a two-temperature model (TTM) \cite{anisimov1974electron} to describe electronic and lattice subsystems' temperature evolution and associated heat transfer. 
To this end, continuum TTM coupled with atomistic MD provides the established route \cite{ivanov2003combined} to simulate both the short-time (e.g., electron excitation, electron-phonon equilibration) and long-time effects (e.g., heating, melting, solidification, recrystallization, phase transformation) encountered in laser-metal interaction within the adequacy of the interatomic potential and computational limit. 
Herein, we refer to this hybrid approach as TTM-MD for simulating ultrafast laser-metal interactions in a physically rigorous way.
\cite{rethfeld2017modelling} provides an extensive review on modeling laser-metal interaction.

Several numerical and experimental studies on ultrafast laser-metal interactions focused solely on bulk configurations \cite{ivanov2003combined, bruneau2005ultra, zhigilei2009atomistic, shugaev2016fundamentals, sokolowski1998transient,siwick2003atomic,sokolowski2003femtosecond,widmann2004single}. 
Some of the earliest atomistic simulation works were dedicated to ablation mechanisms \cite{lewis2009laser,xiong2017effect}.
Even the limited existing laser-thin film works\cite{upadhyay2007response,lin2008electron, gallais2014ultrafast, arefev2022kinetics} did not account for underlying crystallographic orientations and microstructure configurations. 
Consequently, these studies argued the observations predominantly from the energetics standpoint, thus devoid of the microstructure aspects and their crucial interplay.
To this end, the influence of the underlying thin films'  microstructure and crystallographic orientations on the laser-induced deformation remains unexplored, thus least understood. 

Our work addresses this research question by unraveling thin films' crystallographic orientations and microstructure effects/configurations during ultrafast laser-metal interactions using hybrid TTM-MD simulations.
A spatiotemporal variation of absorbed laser energy typically exists depending on local crystallographic and microstructure configurations. 
To account for these variations, we considered a set of applied laser fluences for thin film models with different microstructure and crystallographic orientations. Our simulation results provide a comprehensive understanding of the nanomechanical and thermodynamic aspects of ultrafast laser-metal interactions at atomic resolution, which was previously missing. 

This article is structured as follows: Section \ref{sec:Methods} details hybrid TTM-MD simulations, atomistic modeling of thin films with distinct crystallographic orientations and microstructure topology, interatomic potential, and simulation protocol of ultrafast laser-metal interaction. Section \ref{sec:Results_Discussions} presents insights and critical discussions on the atomic structure evolution, nanomechanics, thermodynamics, and phase transformations of different Au thin films irradiated by a set of applied laser fluences.

\section{Materials and Methods}\label{sec:Methods}

\textbf{\textit{TTM-MD:}} 
In metals, the conduction band electrons absorb the deposited laser energy and undergo collisions with surrounding lattice ions \cite{shugaev2021laser}. The phonons describe the collective behavior of lattice ions within the crystal and play a crucial role in determining electronic and thermal conductivity.
By definition, \textit{classical} MD accounts only for atomic dynamics but not electronic effects (i.e., energy absorption by electrons, subsequent thermalization, and heat transfer from the electrons to lattice ions). 
To include such electronic contributions, a hybrid TTM-MD is often used \cite{ivanov2003combined, leveugle2004photomechanical}, which combines a continuum description of electronic subsystem based on the well-established TTM \cite{anisimov1974electron} and the \textit{classical} MD describing the dynamics of lattice/atomic subsystem.  
More detailed information on TTM-MD can be found, e.g., in \cite{ivanov2003combined, norman2012atomistic,iabbaden2022molecular}.

\begin{figure}
\centering
\includegraphics[scale=.59,trim={10cm 4.7cm 2cm 3cm},clip=true]{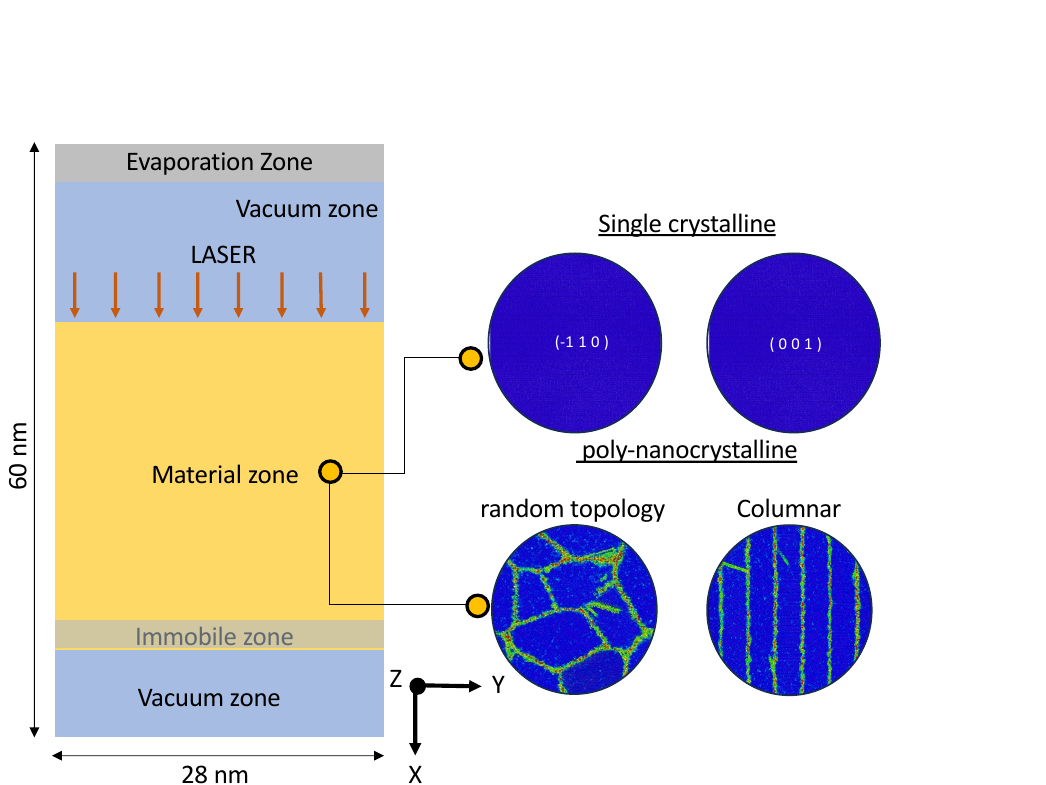}
\caption{Schematic of the atomistic model setup irradiated by the ultrafast laser pulse. The material zone represents the Au thin film, assuming a single crystalline or poly-nanocrystalline configuration. \label{Fig:AtomisticModel}}
\end{figure}  

\textbf{\textit{Atomistic Modeling:}}
Fig. \ref{Fig:AtomisticModel} shows the atomistic thin-film simulation model, comprising the material zone sandwiched between vacuum zones (top and bottom). 
Herein, we considered Au thin films (i.e., material zone) for two specific reasons: 1.) Unraveling our previous experimental study \cite{olbrich2020hydrodynamic} dedicated to single-pulse ultrafast irradiation of Au thin films, 
2.) The assumption of TTM holds well for Au owing to weak electron-phonon coupling \cite{ivanov2003combined}; thus, a longer electron-ion equilibration time \cite{norman2012atomistic}.  
At the bottom material/vacuum interface, we assigned an immobile zone constraining the encompassing atoms' mobility to mimic the substrate's physical constraint \cite{olbrich2020hydrodynamic}.
Atoms leaving the thin films were captured in the evaporation/deletion zone \cite{an2024exploring}. Herein, the ultrafast single-pulse laser irradiates the whole top surface of the material zone indicated by red arrow markers.

The study aims to understand the contributions of and the interplay between the thin film's crystallographic orientations and microstructure topology under ultrafast laser-metal interaction.
Especially with the advent of microstructure-informed atomistic models (MIAMs) \cite{ganesan2021understanding,chandran2024studying}, exploring microstructure-property relationships is nowadays well possible.
To this end, we realized four different Au thin film atomistic models, i.e., two single crystalline (SC) and two poly-nanocrystalline (poly-NC) inspired after MIAMs, all following the general schematic as shown in Fig. \ref{Fig:AtomisticModel}.
For SC, we considered two crystallographic orientation \hkl(-1 1 0) and \hkl(0 0 1), representing different preferred crystallographic orientations during the thin-film growth. 
Likewise, for poly-NC, two different microstructure topologies were considered, namely \textit{random} grain distributions as observed in our previous work \cite{olbrich2020hydrodynamic} and \textit{columnar} grain structures that are often observed in thin film fabrication \cite{rafaja2004interference, karoutsos2012microstructural}.
Furthermore, poly-NC configurations are encountered in thin films due to ultrafast cooling and subsequent solidification from laser irradiation history \cite{olbrich2020hydrodynamic}.

Accordingly, the \textit{random} poly-NC model encompasses 24 arbitrarily shaped grains with an average grain size of ($\approx$ 8.65$\thinspace\si{\nano\metre}$). The other poly-NC model contained 24 \textit{columnar} grains, each with nearly identical hexagonal-shaped with a height of 32$\thinspace\si{\nano\metre}$ (along X) and average face dimensions of 5.8$\thinspace\si{\nano\metre}$ (normal to X).
Our previous work \cite{olbrich2020hydrodynamic} observed an average initial grain size of 50$\thinspace\si{\nano\metre}$. However, our atomistic models considered a smaller grain size and film thickness (i.e., 32$\thinspace\si{\nano\metre}$) to keep the computational load tractable. 
All four atomistic models have the identical box dimensions of 60$\thinspace\si{\nano\metre}$ $\times$ 28$\thinspace\si{\nano\metre}$ $\times$ 28$\thinspace\si{\nano\metre}$ with $\approx$ 1.6 $\times 10^{6}$ Au atoms. 
The crystalline configuration of Au takes a face-centered cubic (fcc) structure with \textit{a} = 4.08\thinspace\AA.

\textbf{\textit{Interatomic potential:}} The electronic configuration of the Au has both delocalized \textit{s} and localized \textit{d} electrons. The former depends only on the electron temperature, whereas the latter also strongly depends on ion coordinates. \cite{mazevet2005ab} reported that a change in electron temperature formally modifies the screening effects of the ions and influences other critical properties (e.g., electron-phonon coupling constant, melting point\footnote{Norman potential predicts reliable melting temperature ($\approx 1336$\thinspace\si{\kelvin}) \cite{mazevet2005ab, arefev2022kinetics}) besides its dependence on the electronic subsystem.}), thus emphasizing the need for electron temperature-dependent interatomic potential for systems where the contribution of localized electrons become important.
The choice of the interatomic potential plays a crucial role in predicting both the ablation mechanisms and corresponding thresholds \cite{zhakhovskii2009molecular}.
Therefore, in this work, we considered such an electron temperature-dependent potential for Au after Norman et al. \cite{norman2012atomistic}, accounting for the electron temperature dependence\footnote{A uniform electron temperature is to be expected for Au thin films with a thickness (i.e., $\leq 100\thinspace\si{\nano\metre}$ for Au) smaller than ballistic energy transport before lattice heating \cite{hohlfeld2000electron}.} on the physical properties of ions, thus faithfully describing the laser-induced processes at atomic resolution, as also demonstrated in some previous works \cite{norman2012atomistic, yao2022exploring}.

The hybrid TTM-MD scheme has a continuum part describing the electron temperature evolution and an atomic part describing the ion dynamics (with modified Newton's equation of motion), including a coupling term for the energy exchange between the two subsystems \cite{rutherford2007effect}.  Technically, the continuum part is realized as spatially resolved grids overlaying the atomistic model. We considered a spatial resolution of 120 $\times$ 8 $\times$ 8 along the X, Y, and Z directions of the whole atomistic model.

\textbf{\textit{Simulation Protocol:}}
In this study, we performed a three-stage simulation protocol as follows:
\begin{enumerate}
    \item \textit{Initial Structure Equilibration} (Stage 1): - global minimization and thermal equilibration of the atomistic models at T = 300$\thinspace\si{\kelvin}$ and P = 1$\thinspace\si{\bar}$, using \textit{classical} MD.
    \item \textit{Laser irradiation} (Stage 2): - simulating the short-time effects, namely electron excitation and subsequent heat transfer from the electronic to the lattice subsystem, using hybrid TTM-MD (for 30\thinspace$\si{\pico\second}$)
    \item \textit{Microstructure \& Thermodynamic evolution} (Stage 3): - simulating the long-time effects, namely heat transport, melting, and phase transformation, using \textit{classical} MD (for another 70\thinspace$\si{\pico\second}$) under NVE conditions.
\end{enumerate}

Following a systematic parametric study, we shortlisted four applied laser fluences (i.e., 8.01\thinspace$\si{\joule/\centi\meter}^2$, 4.80\thinspace$\si{\joule/\centi\meter}^2$, 1.60\thinspace$\si{\joule/\centi\meter}^2$, and 1.28\thinspace$\si{\joule/\centi\meter}^2$) representing high, moderate and low applied fluence, besides considering the target physics (i.e., melting, and ablation) and computational cost. 
Accordingly, in stage 2, we irradiated the surfaces nearly homogeneously\footnote{The dimensions of laser irradiated zone were proportionally reduced to \si{\nano\meter} considering the thin-film model dimension and the computational workload, whereas \si{\micro\meter} was typically reported in experiments \cite{amin2013effect,zhang2014microstructural}.} for all the four atomistic models (i.e., SC\thinspace\hkl(-1 1 0), SC\thinspace\hkl(0 0 1), poly-NC (\textit{columnar}), poly-NC (\textit{random})).
For irradiation, we applied a single-pulse laser beam with the following conditions: 1.) pulse duration, $\tau$ = 100\thinspace\si{\femto\second}, 2.) wavelength, $\lambda$ = 400\thinspace\AA, and 3.) laser penetration depth, \textit{l}$_{skin}$ = 20\thinspace\AA. 

We rationalized such temporal split of hybrid TTM-MD\footnote{chosen after the electron-phonon relaxation time of around 20-30\thinspace\si{\pico\second} both from simulations \cite{ivanov2003combined,lin2008electron} and experiments by measuring the reflectance of the vacuum/gold interface \cite{olbrich2020hydrodynamic}.} (stage 2 for 30\thinspace\si{\pico\second}) and MD\footnote{We limited the total simulation time to 100\thinspace\si{\pico\second} here owing to the focus on laser-induced heating, melting, and ablation.} 
(stage 3 for 70\thinspace\si{\pico\second}) based on the electron-ion equilibration time \cite{norman2012atomistic, ivanov2003combined} and our objective to simulate the interplay of laser-induced perturbation and underlying thin film's crystallographic orientations/microstructure.
For all the stages, we assumed an MD timestep of 1\thinspace\si{\femto\second}.
Such multi-stage simulation protocols were also previously reported \cite{norman2012atomistic, iabbaden2022molecular}.
We assumed full periodic boundary conditions for stage 1. In contrast, for stages 2 and 3, we assumed periodic boundary conditions along Y and Z directions and free boundary conditions normal to the laser irradiated surface in line with several laser-metal studies \cite{ivanov2003combined, iabbaden2022molecular, yao2022exploring}.

\textbf{\textit{Simulation software \& tools:}}
In this work, we realized all \textit{classical} MD and hybrid TTM-MD simulations using the open-source MD code \textit{Large-scale Atomic Molecular and Massively Parallel Simulator}-LAMMPS (stable release version- August 2023) \cite{thompson2022lammps}, distributed by the Sandia National Laboratories. 
For atomistic model construction, we used Atomsk \cite{hirel2015atomsk} and in-house scripts. For visualization and post-processing, we used OVITO \cite{stukowski2009visualization}.
 
\section{Results and Discussions}\label{sec:Results_Discussions}

This section comprehensively discusses the atomic structure evolution, nanomechanics, phase transformations, and thermodynamics of both SC and poly-NC Au thin films irradiated by a set of ultrafast applied laser fluences using TTM-MD (i.e., Stage 2) followed by MD simulations (Stage 3) protocol, as discussed in section \ref{sec:Methods}. 

\subsection{Atomic structure evolution of thin films models:}

\subsubsection{Lattice distortions due to ultrafast laser fluence:}\label{sec:latticeDistortion}

\begin{figure*}[H]
 \centering
 \includegraphics[scale=.75,trim={2cm 3.5cm 2cm 2cm},clip=true]{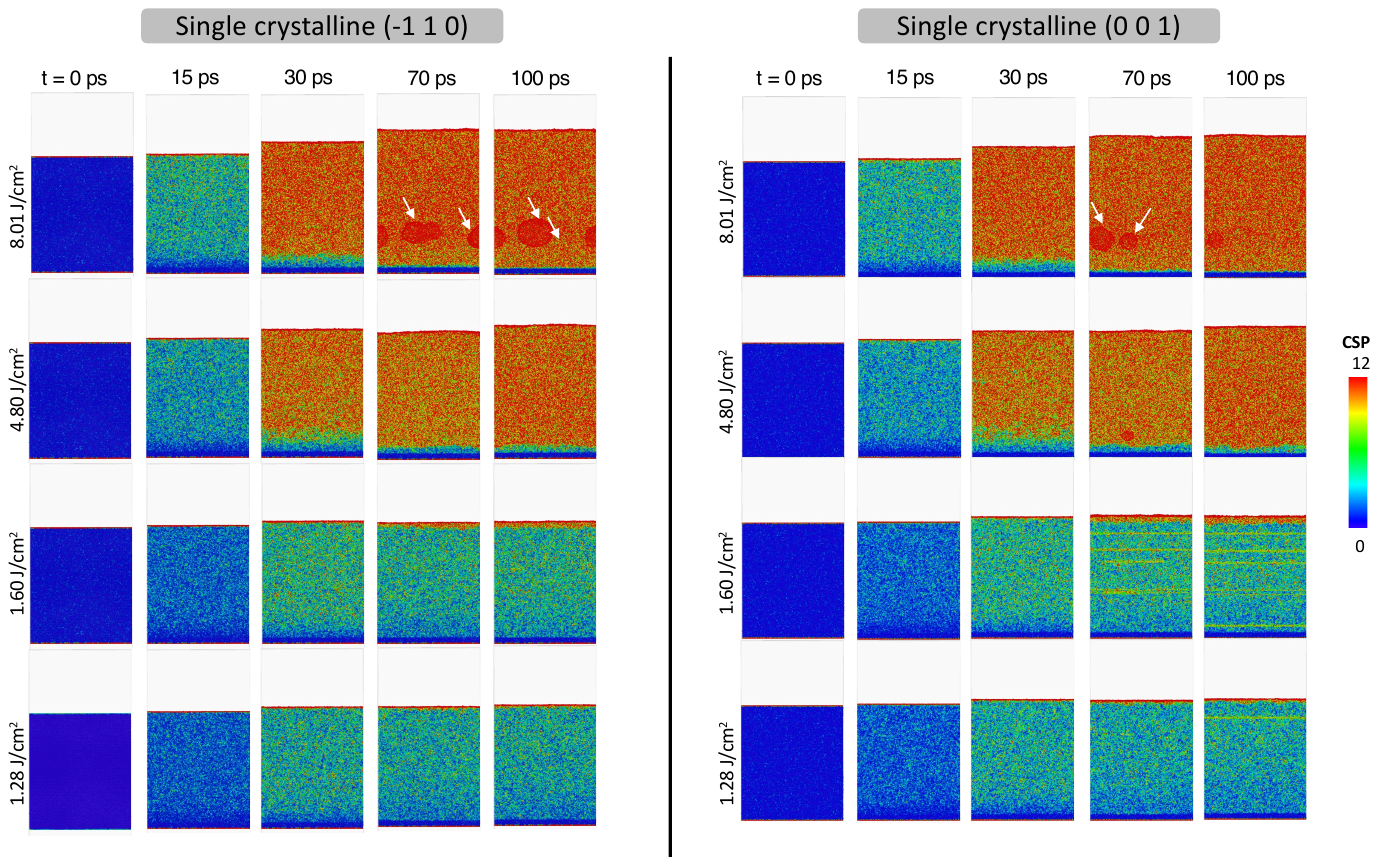}
 \includegraphics[scale=.75,trim={2cm 3.5cm 2cm 3cm},clip=true]{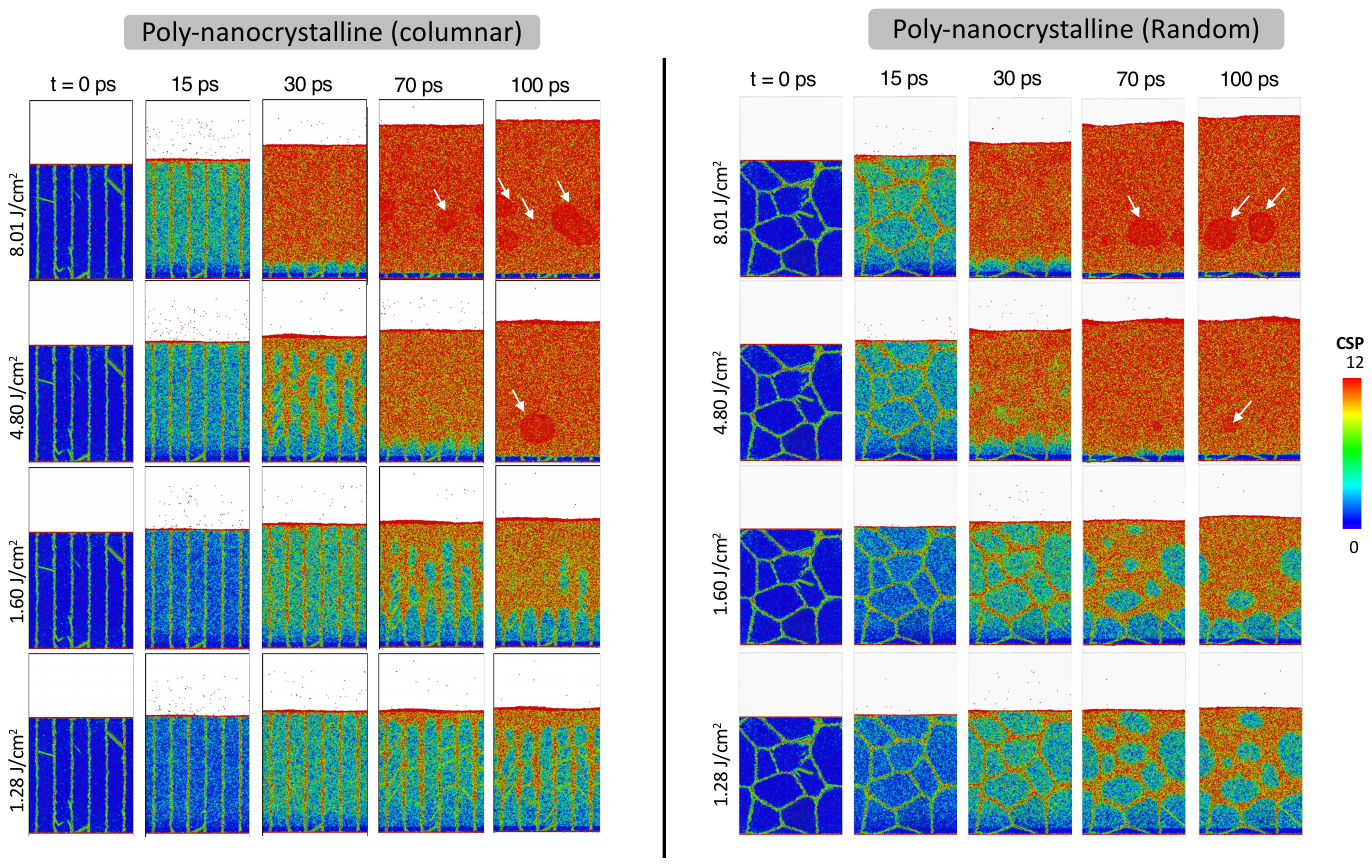}
 \caption{Comparison of single crystalline and poly-nanocrystalline microstructure models irradiated with different applied laser fluence. Atoms are color-coded after centro-symmetry parameter (CSP).}\label{fig:LaserFluenceComparison}
\end{figure*}  

Fig. \ref{fig:LaserFluenceComparison} compares the atomic structure at selected time steps of four thin film models (i.e., SC\thinspace\hkl(-1 1 0), SC\thinspace\hkl(0 0 1), poly-NC (Columnar), and poly-NC (Random)) for different applied laser fluences, respectively. 
Here, we color-coded the atoms after centrosymmetry parameters (CSP), which measure each atom's local lattice distortion by comparing its neighborhood to an ideal crystal reference structure (i.e., Au fcc). 
To this end, unperturbed atoms have lower CSP values close to zero. On the other hand, atoms undergoing lattice distortion deviate significantly from the reference, thus possessing high CSP values.
At 0\thinspace\si{\pico\second}, both SC thin films showed mostly zero CSP (\textit{blue}) as they were initially defect-free except for the surface/interface layer (\textit{red}).
In contrast, at 0\thinspace\si{\pico\second}, both poly-NC thin films exhibited grain boundaries (\textit{green}) with moderate CSP values in addition to defect-free (\textit{blue}) and surface regions (\textit{red}).  
Herein, we used CSP as an effective parameter to observe/track lattice distortion and characterize defects in laser-irradiated Au thin films at atomic resolution. 

\textbf{SC thin films \hkl(-1 1 0) and \hkl(0 0 1):} For the applied laser fluence of 8.01\thinspace$\si{\joule/\centi\meter}^2$ and 4.80\thinspace$\si{\joule/\centi\meter}^2$, we observed severe lattice distortions (i.e., complete melting) of the thin-film region characterized by high CSP values. In particular, the former showed cavitation effects (see the arrow/bubble marker in Fig. \ref{fig:LaserFluenceComparison}). Such complete collapse of crystalline structure in the entire thin-film region suggests homogeneous melting, whereas, for irradiation with moderate (1.60\thinspace$\si{\joule/\centi\meter}^2$) and low applied fluences (1.28\thinspace$\si{\joule/\centi\meter}^2$), no such cavitations were seen, and the lattice distortion/melting confined mainly near the surface, implying the absorbed fluence was below the ablation threshold. 
Such liquid/disordered film region near the surface propagates deep into the film with time, indicating heterogeneous melting.
Interestingly, the SC\hkl(0 0 1) showed the onset of stacking faults traced by dislocations. 
Both SC thin films expanded over time proportional to the applied laser fluences. 

\textbf{poly-NC thin films:} For the applied laser fluence of 8.01\thinspace$\si{\joule/\centi\meter}^2$ and 4.80\thinspace$\si{\joule/\centi\meter}^2$, both the poly-NC models (i.e., Random and Columnar topology) showed the nearly complete collapse of the initial fcc structure or melting of the thin-film region around 100\thinspace\si{\pico\second} with some cavitation bubbles near the bottom. 
In contrast to the SC models, herein, we observed heterogeneous melting characterized by the origin and extent of lattice distortion or melting near the grain boundaries, evidently for both 1.60\thinspace$\si{\joule/\centi\meter}^2$ and 1.28\thinspace$\si{\joule/\centi\meter}^2$.
Under such heterogeneous melting, some crystalline regions (grains) underwent relatively high melting/lattice distortion over others.
Unlike the columnar model, the Au thin film with random grain topology  showed relatively high laser-induced deformation, characterized by extensive grain boundary melting, even at moderate and low laser fluence.
Altogether, poly-NC thin film models underwent large expansion with non-uniform surface topology. 

\textbf{Discussions:}

Our simulations revealed that laser-induced lattice distortions reduced crystal stability and acted as a precursor to melting.
Herein, we limited the focus to melting and ablation\footnote{as there is no phase explosion characterized by rapid melting within a few ps unlike \cite{arefev2022kinetics}} scenarios. 
To this end, for higher applied laser fluences, we observed homogeneous nucleation of the liquid phase (i.e., heavily distorted or melted regions characterized by \textit{red} atoms in Fig. \ref{fig:LaserFluenceComparison}), whereas the lower applied fluences resulted in heterogeneous melting as the solid/liquid interface from the surface propagated deeper into thin films over time.
The magnitude of applied laser fluences decided the melting mechanisms (i.e., homogeneous/heterogeneous).
However, the underlying microstructure topology and local crystallographic orientations seemed to control the rate and extent of melting besides the applied laser fluence.
Consequently, poly-NC films showed comparably more distortions owing to grain-boundary driven melting and expansion than SC thin films (refer to 1.28$\thinspace\si{\joule/\centi\meter}^2$ in Fig. \ref{fig:LaserFluenceComparison}).

We justify this argument based on two factors:

\begin{enumerate}
    \item \textit{Energy}: The applied laser fluences determine the intensity of energy deposition and, thus, the availability of high driving force or activation energy.
    \item \textit{Structure}: Underlying thin-film crystallographic orientations and microstructures determine how the deposited energy gets utilized.    
\end{enumerate}

In the case of SC thin films, the former factor played a dominant role because the laser-deposited (thermal) energy was utilized to overcome the cohesive energies of the solid system (i.e., effectively distorting the crystal structure). 
On the other hand, the poly-NC thin films encompass abundant defects and grain boundaries, as shown in Fig. \ref{fig:LaserFluenceComparison}. Thus, the structural factor in poly-NC films was readily amenable to laser-induced deformation, requiring relatively low energy for mobilizing the available defects. 

Several previous studies \cite{ivanov2003combined, bruneau2005ultra, zhigilei2009atomistic, shugaev2016fundamentals,lewis2009laser,xiong2017effect} attributed the melting mechanisms solely based on the energetic factor (i.e., fluence threshold). 
However, even for the lowest applied laser fluence, poly-NC thin films exhibited relatively high lattice distortion or melting.
Our results suggest that not only the laser-deposited energy (i.e., driving force) but also the underlying thin films' microstructure (i.e., crystallographic orientations and defects) contribute to laser-induced deformation.
To this end, the microstructure features of thin films, namely grain size, grain topology (columnar vs. random), and local crystallographic orientation, seemed to control the rate and extent of lattice disorder evolution or melting under laser irradiation.
Thus, discussing laser ablation only based on the threshold without effectively considering the microstructure aspects and their interplay might not provide a complete picture of ultrafast laser-metal interactions.

\begin{figure*}[H]
\includegraphics[scale=.67,trim={1.8cm 3.5cm 1.5cm 2cm},clip=true]{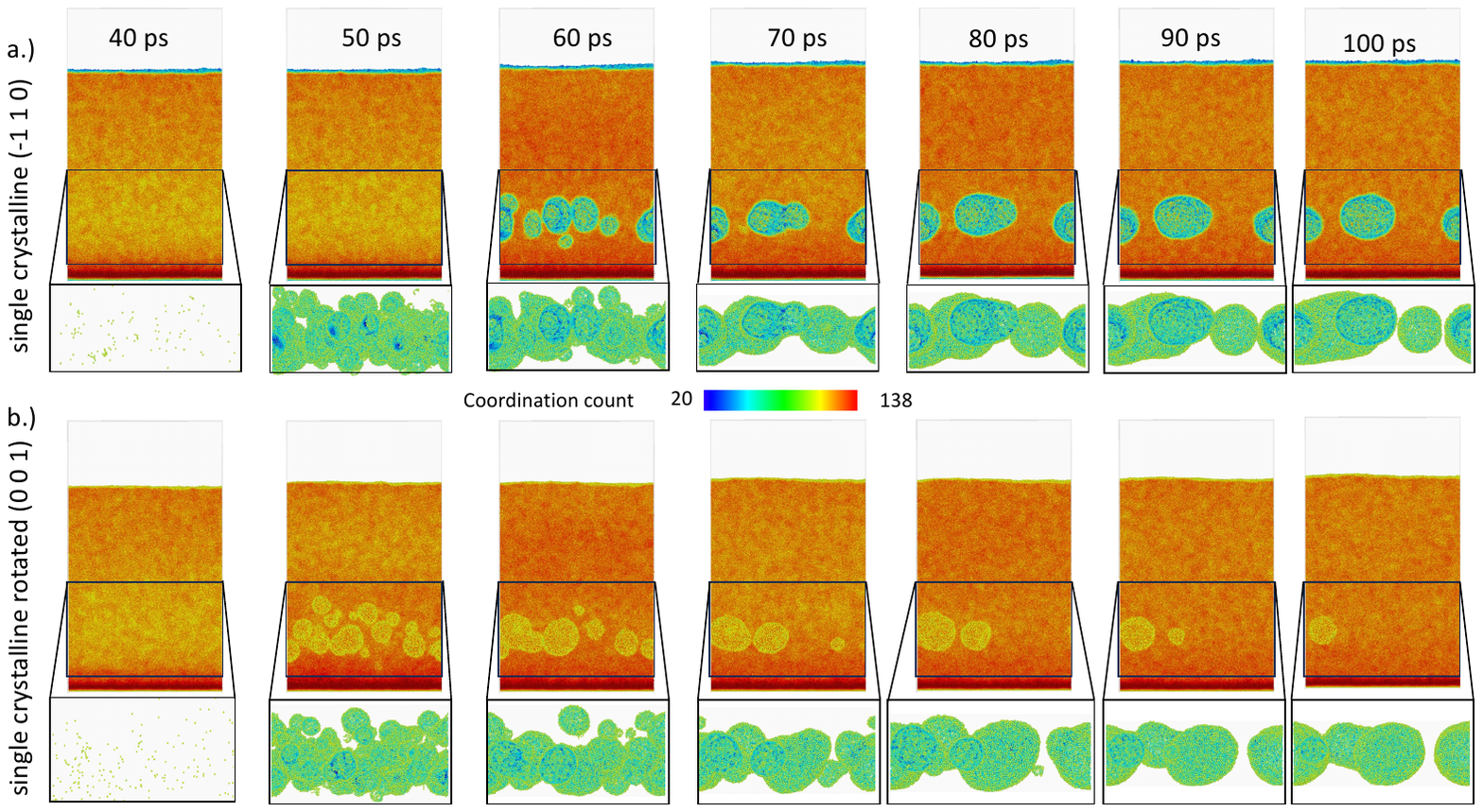}
\includegraphics[scale=.69,trim={1.8cm 3.3cm 1.5cm 2cm},clip=true]{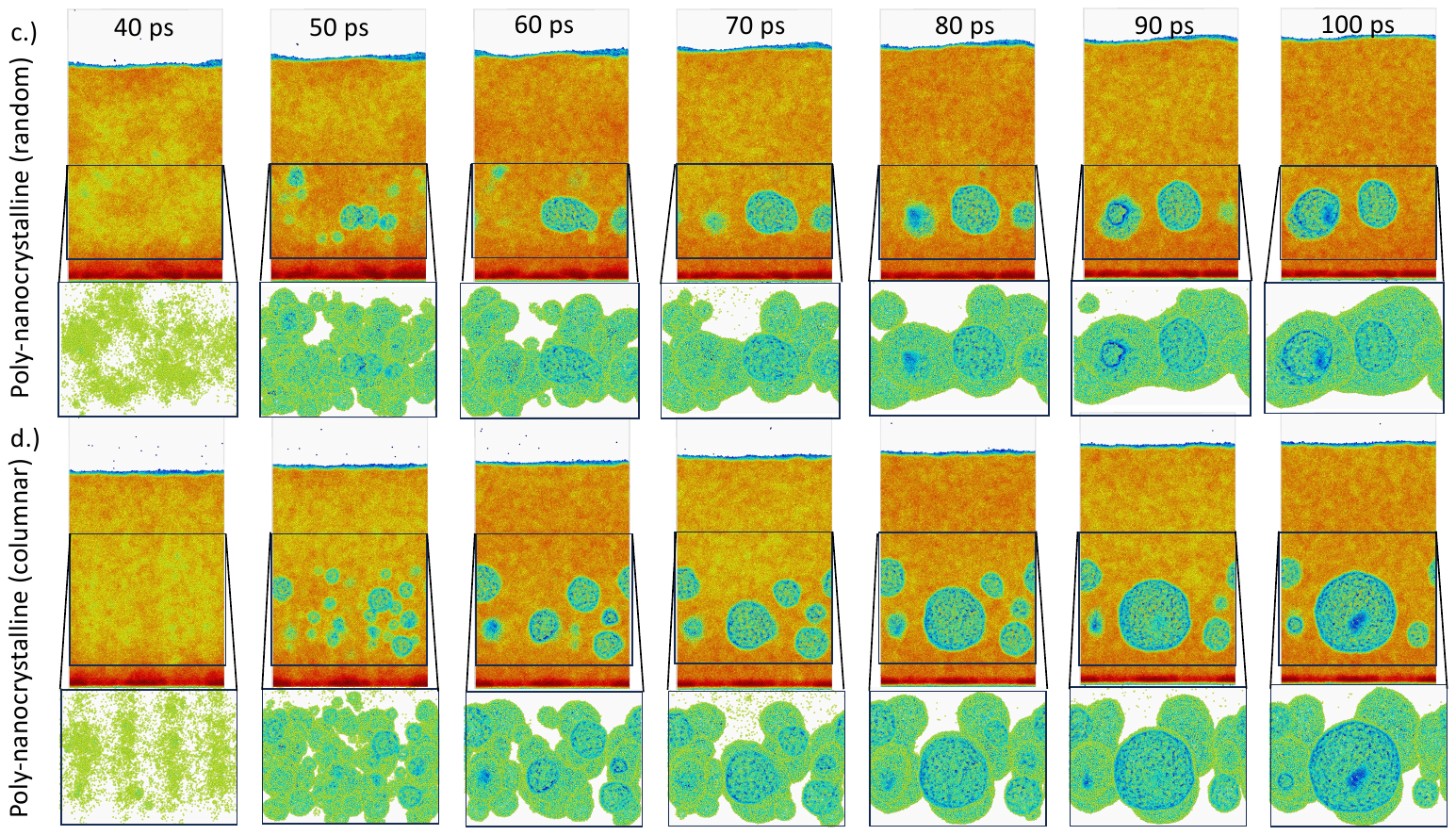}
\caption{Comparison of cavitation effects in different microstructure models under the applied laser fluence of 8.01\thinspace$\si{\joule/\centi\meter}^2$. Cavitations originate due to laser-induced shock wave propagation and resulting thin-film expansion. Herein, atoms with higher coordination counts (> 90) are hidden for clarity to visualize and track cavitation bubbles.\label{Fig:CavitationComparison}}
\end{figure*}  

\subsubsection{Cavitation effects and fragmentation:}
We observed cavitation with subsequent fragmentation for all the thin film microstructure models studied herein at the highest applied fluence of 8.01\thinspace$\si{\joule/\centi\meter}^2$. 
Fig. \ref{Fig:CavitationComparison} shows the origin and evolution of such cavitation for different thin film microstructure models. 
We color-coded the atoms as per the coordination count, where a higher value indicates a solid phase (\textit{high local density}), and a lower one represents a void/surface (\textit{low local density}). 
Due to the likelihood of cavitation, we focused specifically on the mid-to-bottom region of the atomic configuration, and atoms with higher coordination counts (> 90) were hidden for clarity. 

Owing to the electron-phonon coupling, the high deposited laser energy within the ultrafast time (i.e., 100\thinspace\si{\femto\second})  results in rapid lattice heating with a strong temperature gradient, which triggers thermally induced shock waves propagating deep into the thin film. 
Under these conditions, the thin film initially sustained high compression followed by tension, resulting in expansion with the nucleation of several nanovoids.
Consequently, these nanovoids grow and coalesce to form cavitation bubbles. 
All microstructure models experienced cavitation at the highest applied fluence (i.e., 8.01\thinspace$\si{\joule/\centi\meter}^2$), leading to some localized fragmentation of thin films (refer Fig. \ref{Fig:CavitationComparison}).

For both SC thin films (i.e.,\hkl(-1 1 0) \& \hkl(0 0 1)), around 40\thinspace\si{\pico\second}, we detected considerable atoms getting lower coordination count in the lower half portion (refer row 1 \& 2 in Fig. \ref{Fig:CavitationComparison}). 
Then, at 50\thinspace\si{\pico\second}, several nano voids with nearly spherical topology were observed, indicating low local atomic density in the vicinity. 
As the thin film expanded, these nano voids grew in size. They fused with neighboring voids, forming large cavitation channels (with a length of $\approx$ 26\thinspace\si{\nano\meter}) and isolated cavitation bubbles (with a diameter of $\approx$ 7\thinspace\si{\nano\meter}), as shown in the snapshots (e.g., 90\thinspace\si{\pico\second} and 100\thinspace\si{\pico\second}) during melting and expansion (refer Fig. \ref{Fig:CavitationComparison} a and b).

For the poly-NC model (random grains), we observed extensive early onset of nanovoids (characterized by atoms with lower coordination count), even at 40\thinspace\si{\pico\second}, followed by several nanovoids nucleation, resulting from extensive lattice distortion/melting and shock wave propagation.
As the thin film expanded during tension, these nanovoids within the individual and neighboring grains coalesced to form large cavitation channels. In contrast to SC models, the resulting cavitation channels and bubbles were relatively large in dimension with an aspherical topology due to the considerable expansion of the poly-NC thin film.
In particular, for the  poly-NC model (columnar grains), we observed a similar earlier onset of naatoms with lower coordination count; however, spatially distributed along the columnar grain (i.e., along thin-film depth) after 40\thinspace\si{\pico\second} (refer Fig. \ref{Fig:CavitationComparison}). Although the nanovoid nucleation and coalescence mechanisms were comparable, the resulting cavitation topologies were slightly different without any channels. 
These results delineate the contribution of initial thin-film microstructure on ablation and fragmentation profile. 

\begin{figure*}
    \centering
    \includegraphics[scale=.80,trim={4.5cm 4.0cm 0.8cm 2.8cm},clip=true]{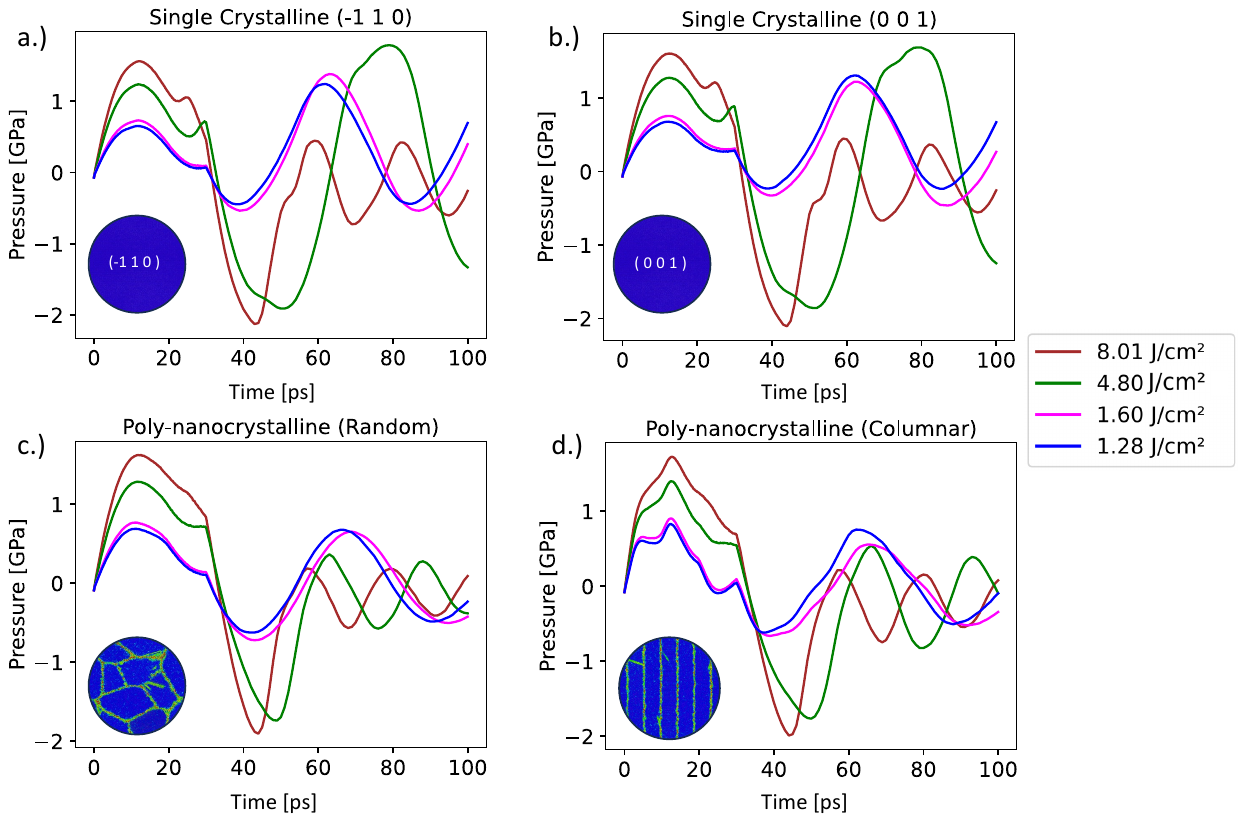} 
    \caption{Evolution of the global pressure computed during stage-2 (TTM-MD) and stage-3 (MD) for four different thin film models: (a) Single Crystalline\thinspace\hkl(-1 1 0) (b) Single Crystalline\thinspace\hkl(0 0 1)  (c) poly-nanocrystalline (Random topology)   (d) poly-nanocrystalline (Columnar).}
    \label{fig:PressurePlots_Global}
\end{figure*}

Herein, the void nucleation events were preceded by a significant increase in global pressure (tension phase), reaching a peak value ($\approx$ 2.1\thinspace\si{\giga\pascal} for SC models vs. $\approx$ 1.8\thinspace\si{\giga\pascal} for poly-NC models) as shown in Fig. \ref{fig:PressurePlots_Global}, after which it dropped towards 0\thinspace\si{\giga\pascal} and remained fluctuating.
As the pressure wave propagated back and forth in the thin-film region, the cavitation bubbles' dimension and topology fluctuated accordingly. 
These pressure fluctuations correspond to laser-induced shock waves.
For lower applied fluences (i.e., < 4.80\thinspace$\si{\joule/\centi\meter}^2$), the SC thin films recorded below the critical pressure ($\approx$ 2.1\thinspace\si{\giga\pascal}), thus, no onset of cavitation bubbles as characterized by fluctuations with high-magnitude (compression and rarefaction in Fig. \ref{fig:PressurePlots_Global}). 
Contrarily, for poly-NC models, we observed cavitation onset for both the applied laser fluences (refer to 8.01\thinspace$\si{\joule/\centi\meter}^2$ and 4.80\thinspace$\si{\joule/\centi\meter}^2$ in Fig. \ref{Fig:CavitationComparison}), with further evidence of dropping pressure towards 0\thinspace\si{\giga\pascal}  (refer Fig. \ref{fig:PressurePlots_Global}). 
Even at moderate/lower laser fluences (i.e., 1.60\thinspace$\si{\joule/\centi\meter}^2$), the fluctuating global pressure for poly-NC thin films recorded lesser peak values than their SC counterparts. 

\textbf{Discussions:}

During the ultrafast laser-metal interaction, one would expect the driving force to be solely thermal energy due to rapid lattice heating.
However, note that we formally introduced two factors: energy and microstructure.
In addition to the thermal energy part, the onset of the laser-induced pressure wave contributes to the thin film's compression and tension (expansion). 
Note that pressure-driven tension/expansion denotes the mechanical energy part, which complements the reduced thermal component \cite{ivanov2003combined} owing to thermomechanical contribution. 

Both from a qualitative (refer Fig. \ref{Fig:CavitationComparison}) and quantitative (refer Fig. \ref{fig:PressurePlots_Global}) standpoint, we would argue that these cavitations act as pressure relieving mechanisms, activated once the laser-induced pressure in thin film reaches a threshold. 
Herein, our results suggest such cavitation thresholds were sensitive to underlying thin-film microstructure topology and crystallographic orientations, shown by lower peak pressure for poly-NC thin films than SC counterparts (refer Fig. \ref{fig:PressurePlots_Global}).
Recently, \cite{protim2023two} reported a  higher concentration of voids in Cu thin films irradiated by a femtosecond laser pulse (high energy transfer) compared to the picosecond laser pulse (moderate energy transfer). Our work delineated how the underlying microstructure topology and crystallographic orientations facilitate higher void concentrations or energy utilization. 
In particular, the cavitation activities resulted in thin film fragmentation for higher applied laser fluences. 
Thus, the underlying microstructure, characterized by the existence and volume fraction of the grain boundaries, influences the thin films' melting, ablation, and fragmentation behavior.

\begin{figure*}
\centering
\includegraphics[scale=.69,trim={1.3cm 3.0cm 1cm 2cm},clip=true]{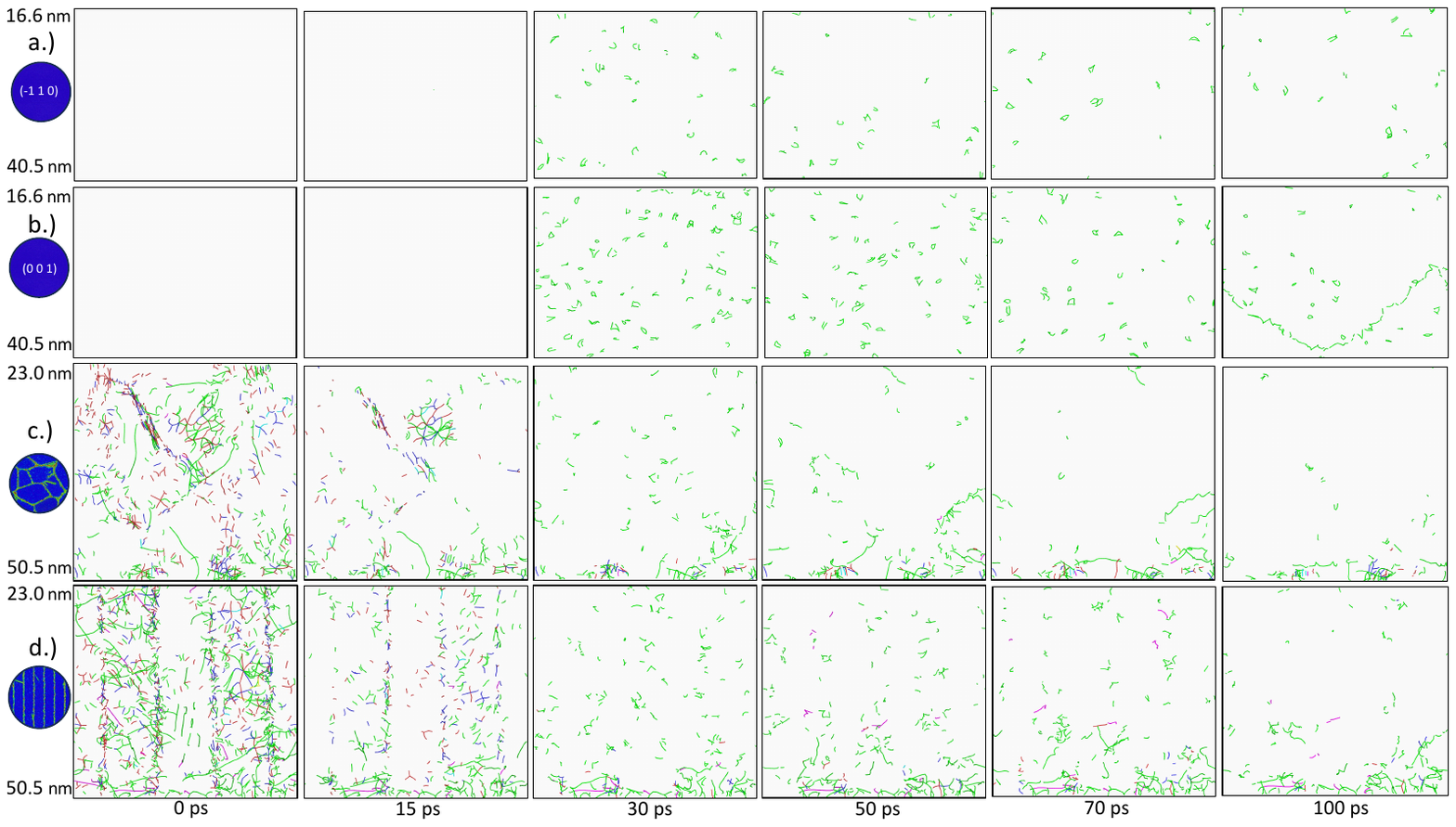}
\caption{Dislocations evolution comparison in underlying Au thin-film microstructure for laser fluence of 1.281\thinspace$\si{\joule/\centi\meter}^2$.\label{Fig:DislocEvolution}}
\end{figure*}  
 
\subsubsection{Laser induced defects in thin films:}

In this subsection, we investigate the nanomechanical aspects of ultrafast laser-metal interaction that still account for thin film's crystallographic orientation and microstructure topology.

Fig. \ref{Fig:DislocEvolution} shows the dislocation evolution along the film depth (\textit{for the selected observation zone}) after \cite{stukowski2009visualization}. 
For this investigation, we selected such lower applied fluence (1.281\thinspace$\si{\joule/\centi\meter}^2$) as the thin films retained most of the initial crystalline structure (refer Fig. \ref{fig:LaserFluenceComparison}), thus aiding defect characterization.
For both SC thin films (refer to rows a, b in Fig. \ref{Fig:DislocEvolution}), we observed underneath the irradiation surface (i.e., from 16.6\thinspace\si{\nano\meter} to 40.5\thinspace\si{\nano\meter}), whereas for poly-NC thin films we observed slightly deeper (i.e., from 23.0\thinspace\si{\nano\meter} to 50.5\thinspace\si{\nano\meter}) considering active dislocations regions (refer to rows c, d in Fig. \ref{Fig:DislocEvolution}).

The initially defect-free SC thin films showed no traces of dislocations until 20\thinspace\si{\pico\second}.
This observation suggests that under such lower applied fluence (1.281\thinspace$\si{\joule/\centi\meter}^2$),
no significant lattice distortion occurred (initially after laser irradiation) except for some thin-film compression and subsequent heating (see also Fig. \ref{fig:LaserFluenceComparison}).   
Around 30\thinspace\si{\pico\second}, we observed nucleation of several dislocation segments. 
This nanomechanical event shall be correlated with the kink on the global pressure curve during the tension stage (refer to Fig. \ref{fig:PressurePlots_Global}).
Subsequently, the dislocations declined owing to laser-induced thin-film expansion (i.e., tension). 
However, for the same applied laser fluence, the SC\hkl(0 0 1) exhibited slightly different spatiotemporal dislocations than SC\hkl(-1 1 0), thus implying the influence of thin film's crystallographic orientation on laser-induced defects.
Characteristically, both SC thin films showed predominantly Shockley partial dislocations.  
Unlike SC models, both poly-NC models (at 0\thinspace\si{\pico\second}) possessed initial grain boundary misfit dislocations inline with the underlying grain topology (i.e., random vs. columnar), as shown in Fig. \ref{Fig:DislocEvolution} (rows c and d). 
Also, poly-NC thin films exhibited relatively high laser-induced deformation for the lowest fluence. 
Herein, we observed qualitatively that as the thin film undergoes melting and phase transformation, more dislocations dissipate, owing to a decline in crystalline configuration.
Consequently, even for such lowest applied fluence in poly-NC thin film, the decline in dislocations validates grain boundary melting. 

\begin{figure*}
    \centering
    \includegraphics[scale=.75,trim={2.0cm 4.5cm 0.8cm 1.8cm},clip=true]{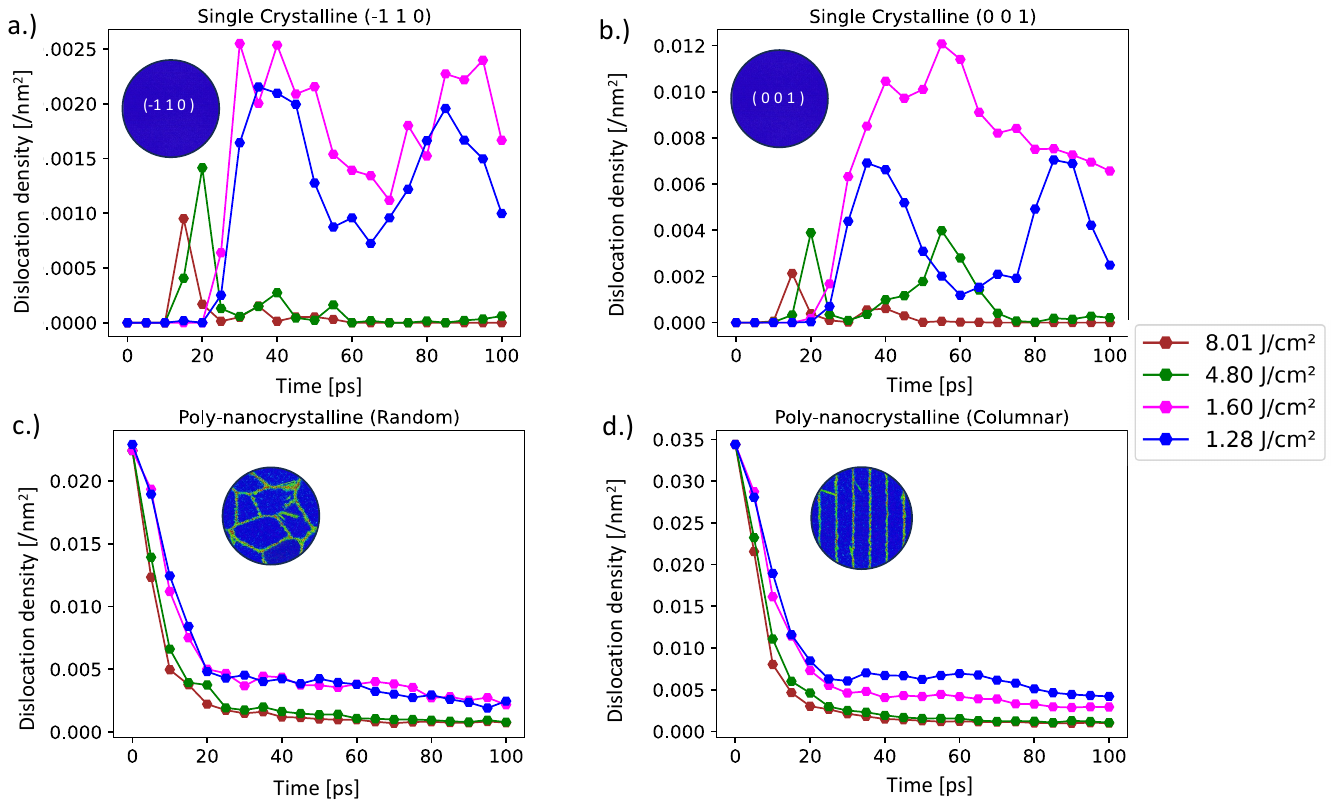} 
    \caption{ Dislocation density evolution in Au thin films for different applied laser fluences for four underlying configurations: (a) Single Crystalline\thinspace\hkl(-1 1 0), (b) Single Crystalline\thinspace\hkl(0 0 1),  (c) Nanocrystalline (Random topology),   (d) Nanocrystalline (Columnar).}
    \label{fig:DislocDensityPlots_Global}
\end{figure*}

Altogether, SC thin films withstood higher laser-induced pressure waves than poly-NC thin films. In the former, the pressure was utilized for defect nucleation and lattice distortion, whereas the latter encompassed preexisting grain boundary dislocations; thus, there was no significant rise in pressure. 
To this end, existing dislocations validate the thin films' crystallographic spatial regions (i.e., solid phase), serving as descriptors of the plastically deformed crystalline phase at any given time.
Qualitatively, the thin film's grain topology, grain size, and crystallographic orientation influenced laser-induced defect evolutions (refer Fig. \ref{Fig:DislocEvolution}). 
However, a quantitative assessment becomes instrumental to unravel the nanomechanics of laser-induced defects in thin films.

Fig. \ref{fig:DislocDensityPlots_Global} shows the  dislocation density ($\rho_{Dis}$) evolution of different thin film models irradiated by a set of applied laser fluences  studied herein:

\textbf{SC thin films:} For higher applied laser fluence (8.01\thinspace$\si{\joule/\centi\meter}^2$), both SC thin films exhibited a short-lived peak in $\rho_{dis}$ after 15\thinspace\si{\pico\second}, indicating dislocation nucleation events and subsequent drop in $\rho_{dis}$ to nearly zero (refer Fig. \ref{fig:DislocDensityPlots_Global} a), indicating, complete melting or collapse of crystalline structure in thin films. 
Then, for the applied fluence (4.80$\thinspace\si{\joule/\centi\meter}^2$), a slightly delayed peak of $\rho_{dis}$ around 20\thinspace\si{\pico\second}, which drops further to nearly zero with negligible occasional peaks. 
However, the SC\thinspace\hkl(0 0 1) showed a second episode of dislocation activity from 30 to 55 \thinspace\si{\pico\second} before dropping again to zero (owing to different activated slip systems). 
Interestingly, for 1.60\thinspace$\si{\joule/\centi\meter}^2$ and 1.28\thinspace$\si{\joule/\centi\meter}^2$, both SC thin films showed an active dislocation generation phase starting from 20\thinspace\si{\pico\second} until 30\thinspace\si{\pico\second}. 
This time window corresponds with the thin film expansion (i.e., the tension stage in Fig. \ref{fig:PressurePlots_Global}).
Subsequently, the short dip shall be attributed to annihilation (\textit{due to dislocation mobility}) and dissipation events (\textit{near solid/liquid interface}) during the compression stage owing to significant lattice distortion.

Generally, the two SC thin films demonstrated qualitatively similar trends in $\rho_{Dis}$ evolution irradiated by the ultrafast laser pulse.
However, quantitatively, SC\thinspace\hkl(-1 1 0) yielded $10^{1}$ less $\rho_{Dis}$ than SC\thinspace\hkl(0 0 1).
This difference shall stem from two aspects:
1.) \textit{Dislocation starvation zone:}
the ease of activated slip systems and associated energy requirements (dependence on local crystallographic orientations).
2.) \textit{Crystalline volume fraction:}
the underlying volume fraction of deformed/available crystalline regions plays a decisive role in thin films.

\begin{figure*}
\centering
\includegraphics[scale=.65,trim={1.3cm 5cm 1cm 2cm},clip=true]{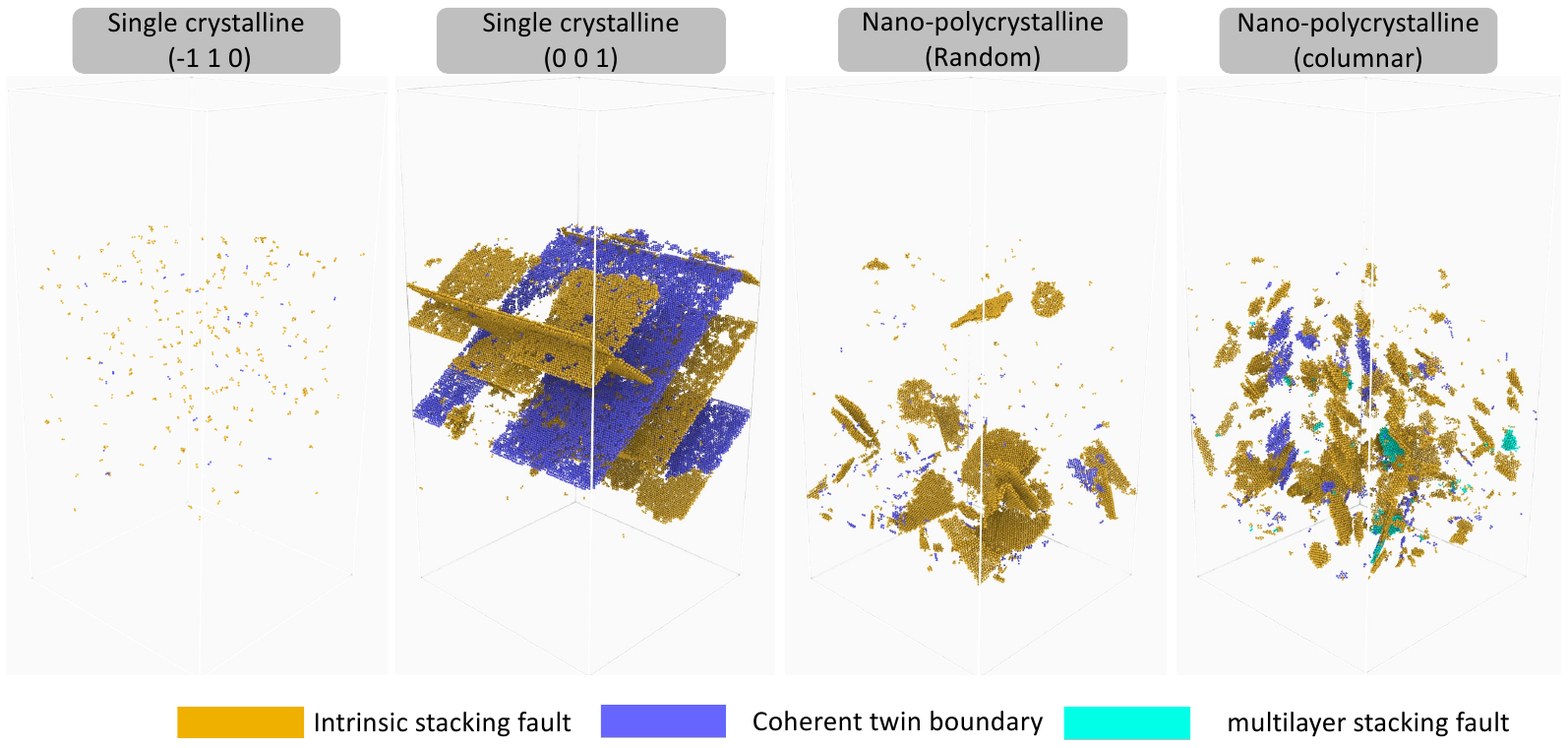}
\caption{Comparison of planar defects for different thin-film models at 70\thinspace\si{\pico\second} under the fluence of 1.281\thinspace$\si{\joule/\centi\meter}^2$.\label{Fig:StackingFault}}
\end{figure*}  

\textbf{poly-NC thin films:} For all the applied laser fluences, both poly-NC thin films exhibited a sharp drop in $\rho_{Dis}$ from the start until 20\thinspace\si{\pico\second} and decay slowly until 100\thinspace\si{\pico\second}.
Such an exponential reduction in $\rho_{Dis}$ was attributed to the rapid/relatively high laser-induced melting or collapse of crystalline structure observed in poly-NC models (refer Fig. \ref{fig:LaserFluenceComparison}) studied herein. However, the slight difference in the residual\footnote{The residual $\rho_{Dis}$ correspond to misfit dislocations at grain boundaries in the immobile region of the atomistic model (refer Fig. \ref{Fig:AtomisticModel}).}
$\rho_{Dis}$ magnitude between high and moderate fluence remained proportional to the extent of melting/loss of crystallinity (i.e., annihilation/dissipation phase). 

\textbf{Probing planar defects in thin films:}

Fig. \ref{Fig:StackingFault} shows the identified planar defects \cite{stukowski2009visualization} (i.e., intrinsic stacking faults (ISF), coherent twin boundary (CTB), multilayer stacking fault (MSF)) in both the laser-irradiated SC and poly-NC thin films at 70\thinspace\si{\pico\second} for 1.28\thinspace$\si{\joule/\centi\meter}^2$.
In contrast to SC\thinspace\hkl(-1 1 0) thin-film, SC\thinspace\hkl(0 0 1) showed extensive planar defects (i.e., ISF and CTB), owing to the activation of favorable slip systems.
On the other hand, the poly-NC thin film (\textit{random} topology) exhibited substantial ISF and relatively low CTB. Some crystalline grains showed disproportionately high SFs over others owing to locally preferred crystallographic orientations, thus influencing activated planar defects. 
The poly-NC thin film (\textit{columnar} topology) exhibited a mixture of planar faults with high ISFs, followed by CTBs and MSFs.

Recall that by construction, the SC thin films possess large grain sizes ($d_{grain}\approx 30\thinspace\si{\nano\meter}$) with one characteristic crystallographic orientation. In contrast, both the poly-NC thin films contained a set of medium-sized grains ($\approx 7\thinspace\si{\nano\meter}$), each with distinct crystallographic orientations (refer sec. \ref{sec:Methods}). 
Besides dislocations, the ultrafast laser irradiation of metallic thin films also induced planar defects as a pressure-relieving mechanism (refer Fig. \ref{Fig:StackingFault}.
We could elucidate such pressure relieving mechanism from the global pressure trend (refer second peaks for 1.28\thinspace$\si{\joule/\centi\meter}^2$ in Fig. \ref{fig:PressurePlots_Global}), where SC\thinspace\hkl(0 0 1) showed relatively lower compression (heating) and higher tension (planar faults/defects nucleation), vs. SC\thinspace\hkl(-1 1 0) showing the higher compression and lower tension (defect annihilation/dissipation), weakly correlating to the planar defects. 

The onset of planar defects as pressure-relieving mechanisms delineate the nanomechanics behind the utilization of deposited thermal energies during thin-film expansion, which is sensitive to crystallographic orientations, grain topology, and grain size. 
In particular, such an intermediate plastic regime (i.e., planar defects) delays thin films' fracture/fragmentation (e.g., void creation).
To this end, dislocations  and planar defects shall be considered primary pressure-relieving mechanisms when the system is crystalline, whereas cavitations were pressure-relieving events upon phase transformation.
The nanomechanical aspects revealed that thin film straining depends on the grain size and preferred crystallographic orientation, accommodating/relieving laser-induced stresses accordingly.
These critical insights could be applied to strengthen the target material and modify the underlying microstructure with minimal damage \cite{titus2015dislocation}.

\begin{figure*}[H]
\centering
\includegraphics[scale=.74,trim={2.0cm 2.9cm 1cm 2cm},clip=true]{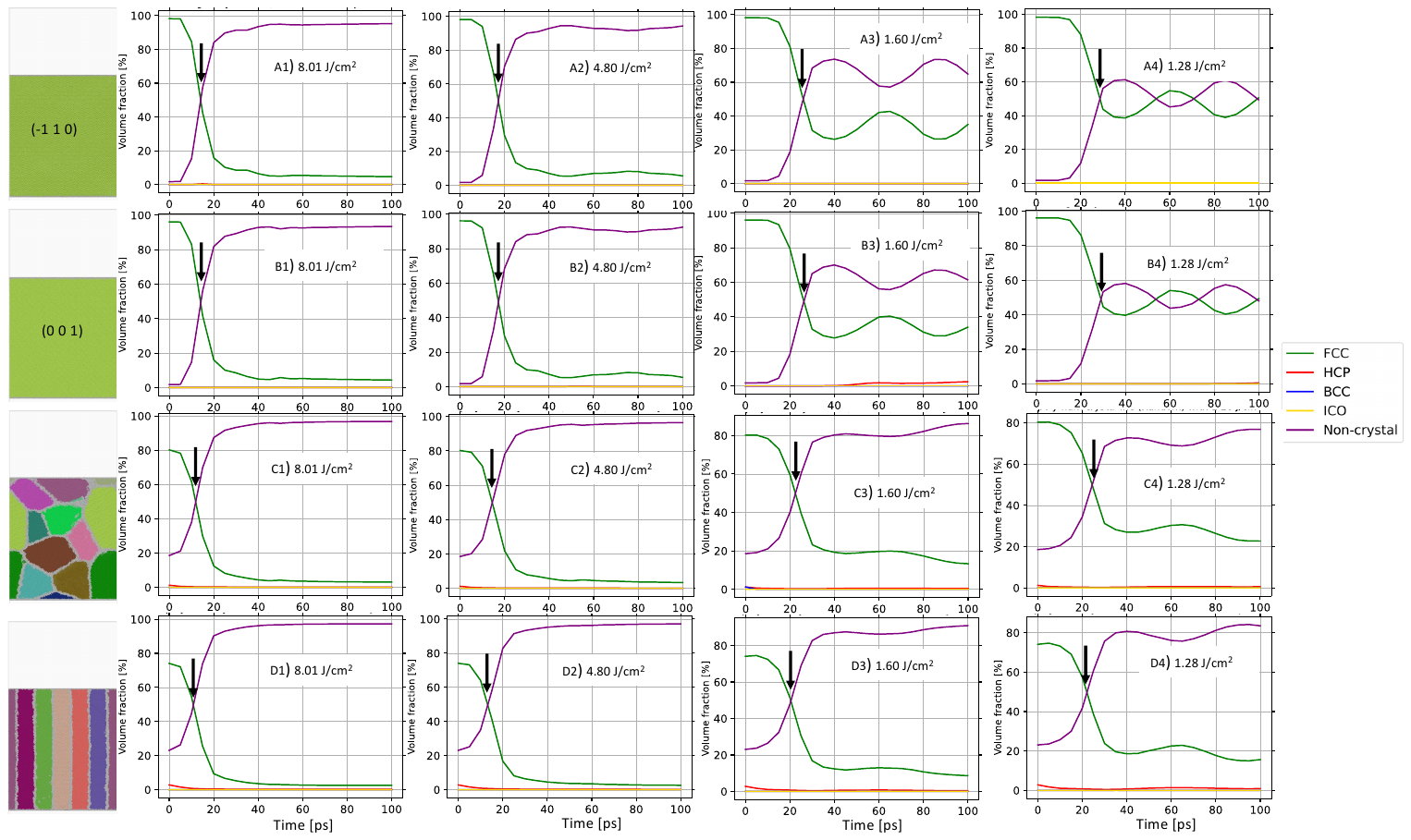}
\caption{Comparison of grain/crystalline regions evolution (volume fraction) for four Au thin-film atomistic models irradiated with different applied laser fluences.\label{Fig:VolumeFraction}}
\end{figure*}  

\subsection{Phase transformation and thermodynamic state evolution:}
\subsubsection{Effects of grain topology and crystallographic orientation:}

The strong correlation between defect density and melting regions delineates the role of underlying thin film microstructure topology and crystallographic orientations. 
Tracking the volume fraction of the crystalline region could shed more light on why some configurations were amenable to enhanced laser ablation over others and how these interesting insights can be applied to optimize nanoscale processing.

Fig.\ref{Fig:VolumeFraction} shows the evolution of crystalline and noncrystalline region volume fraction for SC and poly-NC thin film models under various applied laser fluences. 
Herein, we classified the local structure of each atom's neighborhood using the Polyhedral Template Matching method \cite{larsen2016robust}. 
The initially defect-free SC thin film (i.e., \hkl(-1 1 0) and \hkl(0 0 1)) showed a high starting fcc volume fraction of 98\% and 95\%, respectively (refer A1 and B1 in Figure. \ref{Fig:VolumeFraction}). 
With the onset of melting and subsequent expansion, we observed significant phase transformation (i.e., solid $\rightarrow$ crystalline disorder $\rightarrow$ liquid), characterized by a combination of a steady decline in fcc and a corresponding increase in the noncrystalline region (i.e., \textit{marking the collapse of crystalline structure}).
The crossing points (refer to arrow markers in Fig. \ref{Fig:VolumeFraction} highlight the 50\% fcc volume fraction. 
Here, the crossing points for each thin film model were delayed with decreasing applied laser fluence, indicating a reduced melting rate. 
At higher fluences (8.01\thinspace$\si{\joule/\centi\meter}^2$, 4.80\thinspace$\si{\joule/\centi\meter}^2$), SC thin films after 100\thinspace\si{\pico\second} showed nearly complete phase transformation/melting (i.e., resulting in 5\% and 8\% fcc, respectively).
However, lower fluences (i.e., 1.60\thinspace$\si{\joule/\centi\meter}^2$, 1.28\thinspace$\si{\joule/\centi\meter}^2$) retained some 30\% to 40\% of initial fcc volume fraction with fluctuations owing to traversing pressure waves in thin films. To this end, both SC thin films showed qualitatively similar and consistent phase transformation trends, with minor quantitative differences for lower fluences.

The two initial poly-NC thin films (i.e., random and columnar) exhibited $\approx$ 20\% non-crystalline volume fraction corresponding to grain boundaries. 
Herein, the net phase transformation (i.e., fcc $\rightarrow$ noncrystalline) was comparable across all applied laser fluences studied (i.e., $\approx$ 80\% for higher fluence followed by 60\% for lower fluence).
Unlike SC thin films, no strong fluctuations in fcc volume fraction were observed due to extensive melting/lattice disorder, indicating subsequent loss of crystalline configuration. 
Furthermore, in poly-NC thin films, the preexisting grain boundaries accelerated the phase transformation compared to SC thin films, indicated by earlier crossing points compared to SC thin films (refer to arrow markers in Fig. \ref{Fig:VolumeFraction}). 
These results delineate the extensive early melting/lattice distortions of poly-NC thin films (refer section \ref{sec:latticeDistortion} and Fig.\ref{fig:LaserFluenceComparison}) for each applied laser fluence.
Herein, among the four thin film models investigated, poly-NC (columnar) showed extensive melting characterized by loss of crystallinity.
To this end, the thin film's crystallographic orientation and grain topology influence the extent of laser-induced deformation processes and melting. 

\subsubsection{Effects of pressure and temperature in Au thin films:}

\begin{figure*}[H]
\centering
\includegraphics[scale=.72,trim={1.8cm 5.7cm 1cm 4.5cm},clip=true]{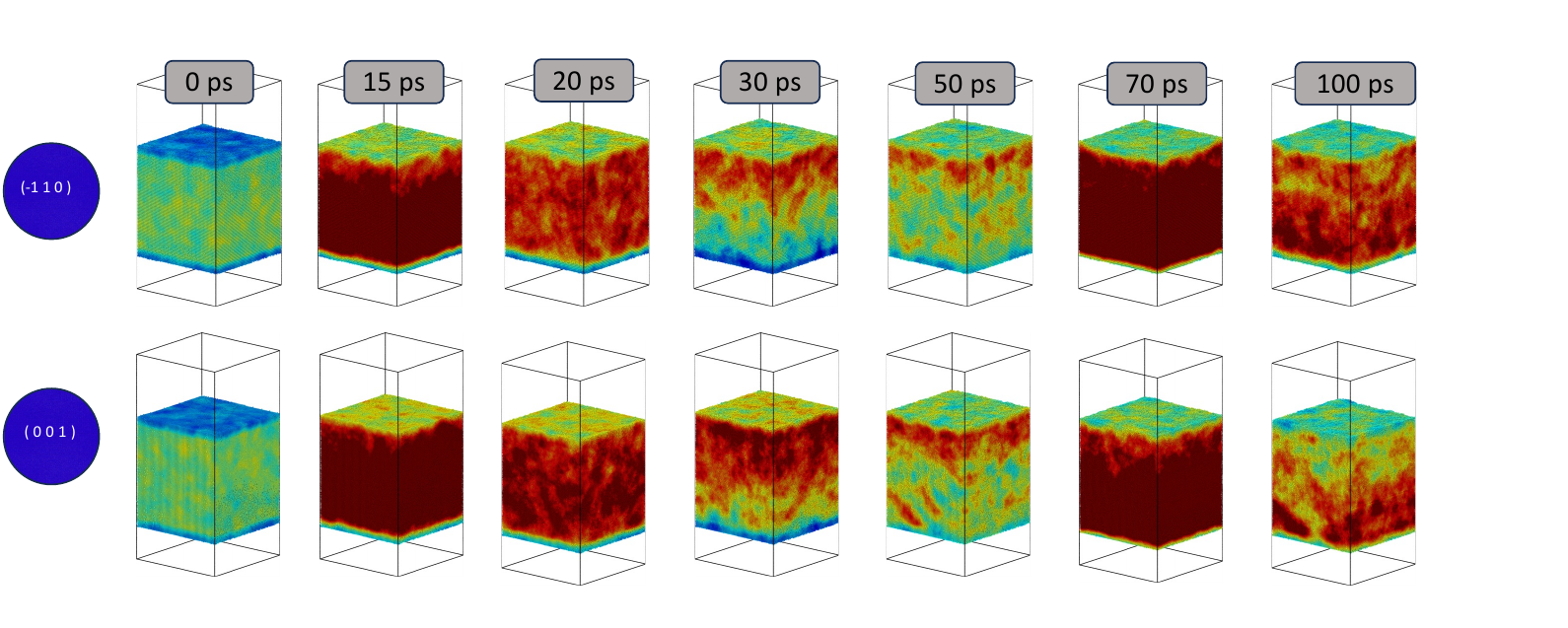}
\includegraphics[scale=.72,trim={1.8cm 4.7cm 1cm 6.3cm},clip=true]{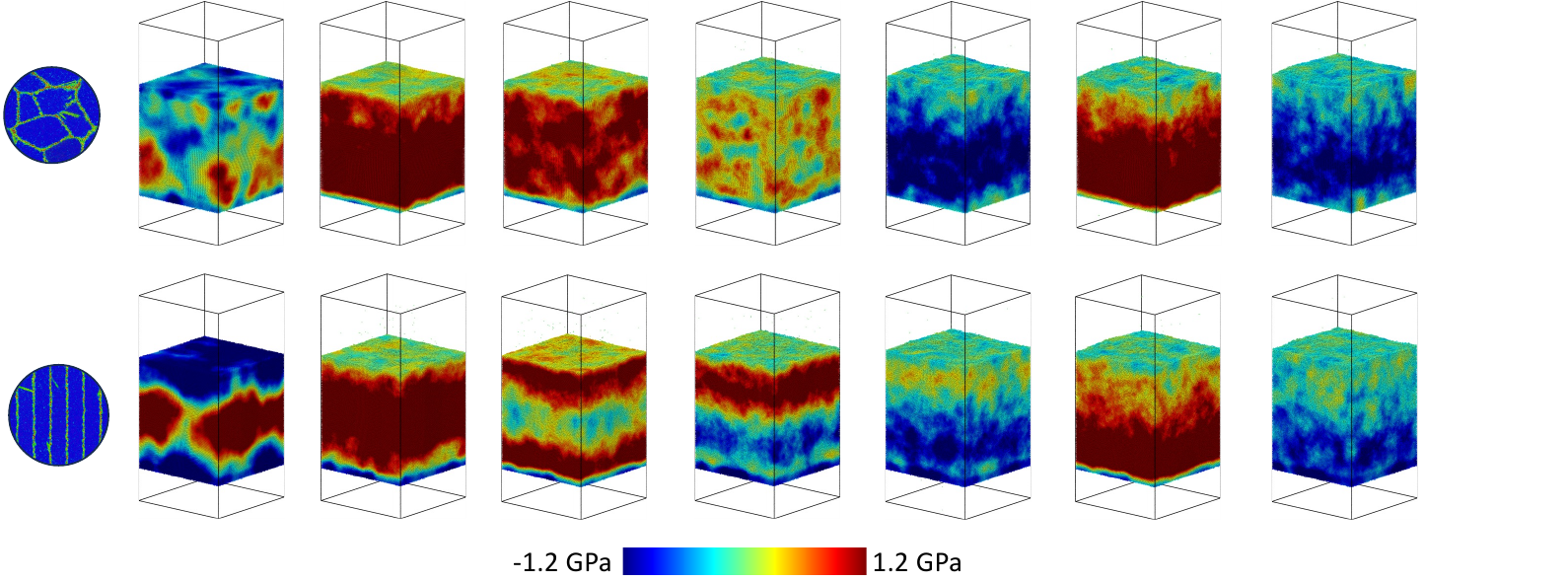}
\caption{Spatio-temporal evolution of laser-induced pressure for different Au thin-film models irradiated with $ 1.28\thinspace\si{\joule/\centi\meter}^2$.}\label{Fig:SpatioTemporal_PressureEvolution}
\end{figure*}  

Fig.\ref{Fig:SpatioTemporal_PressureEvolution} shows the spatiotemporal pressure evolution in thin films under varying crystallographic orientations (SC) and microstructures (poly-NC) at 1.28\thinspace$\si{\joule/\centi\meter}^2$. 

\textbf{Pressure Effects:} The two SC thin films (i.e., \hkl(-1 1 0) \& \hkl(0 0 1)) start from a nearly homogeneous pressure distribution.
A rapid high-energy transfer from the electronic to the lattice subsystem produces a strong temperature gradient, triggering pressure waves. 
At the start, thin films resisted some atomic structural change by undergoing compression with a peak until 15\thinspace\si{\pico\second} (refer to positive values on the color code in Fig. \ref{Fig:SpatioTemporal_PressureEvolution}). 
Subsequently, the thin film expanded, marking the tension stage (i.e., the transition towards negative values on the color code) until 45\thinspace\si{\pico\second}. 
Herein, the time window of the tension stage corresponds well with both global pressure plot (refer Fig. \ref{fig:PressurePlots_Global}) and dislocations nucleation events (refer Fig. \ref{Fig:DislocEvolution}, \ref{fig:DislocDensityPlots_Global}) for 1.28\thinspace$\si{\joule/\centi\meter}^2$.
Owing to the further oscillating pressure wave, SC thin films underwent compression again, reaching new peak values around 70\thinspace$\si{\joule/\centi\meter}^2$. Such a recurring compression state shall be attributed to a combination of factors: a.) due to interactions of reflecting pressure waves \cite{olbrich2020hydrodynamic}, b.) high compressibility of highly disordered/molten Au state, c.) absence of cavitation bubbles, thus, no pressure relieving mechanism, as also shown by global pressure (Fig. \ref{fig:PressurePlots_Global}). 

In contrast to SC thin films, Poly-NC showed an inhomogeneous pressure distribution (even from the start), which could be attributed to spatially varying crystallographic orientations and grain topology (microstructure fingerprints).
Furthermore, these thin films accumulated no strong compressive pressure due to considerable lattice distortion driven by grain boundaries and grain orientation dependent straining.
Such a spatiotemporal distribution of moderate tension (\textit{blue}) and compression regions (\textit{red}) correspond very well with global pressure plots (refer Fig.\ref{fig:PressurePlots_Global}).
These results delineate the role of underlying microstructure topology and crystallographic orientations in determining the pressure distributions and magnitude.

\begin{figure*}[H]
\centering
\includegraphics[scale=.72,trim={1.8cm 6.75cm 1cm 4.5cm},clip=true]{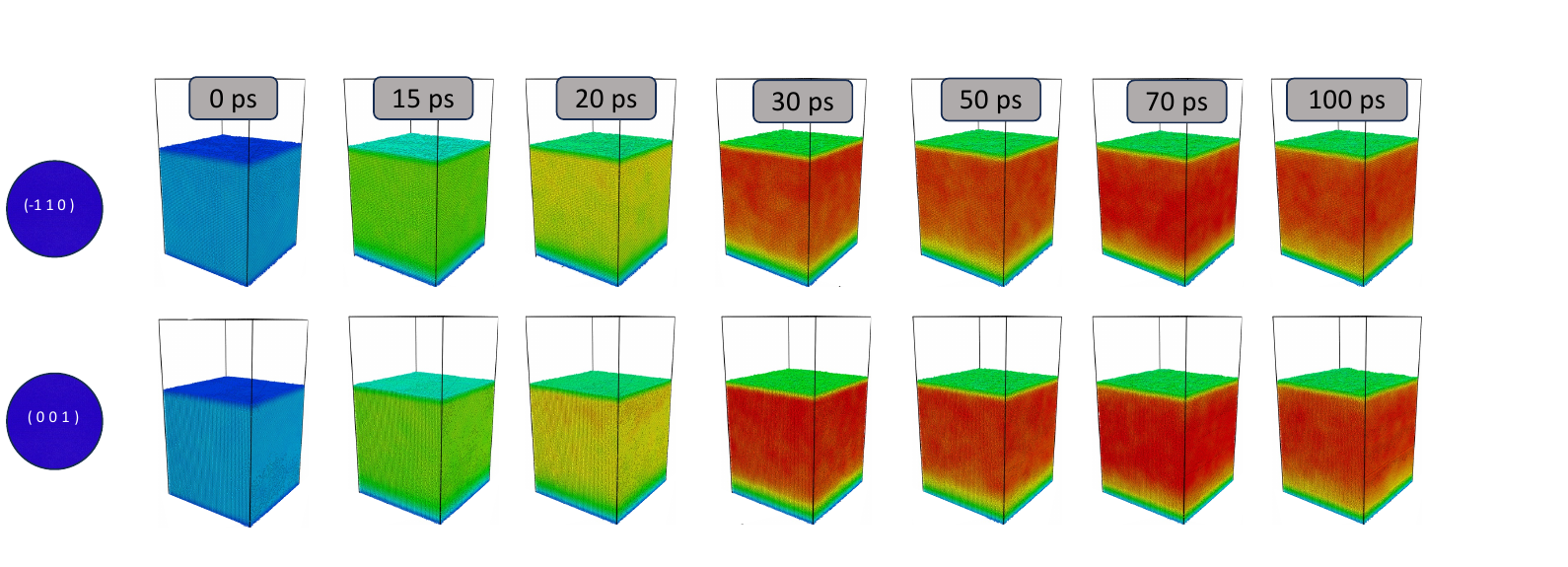}
\includegraphics[scale=.72,trim={1.8cm 5.9cm 1cm 6.2cm},clip=true]{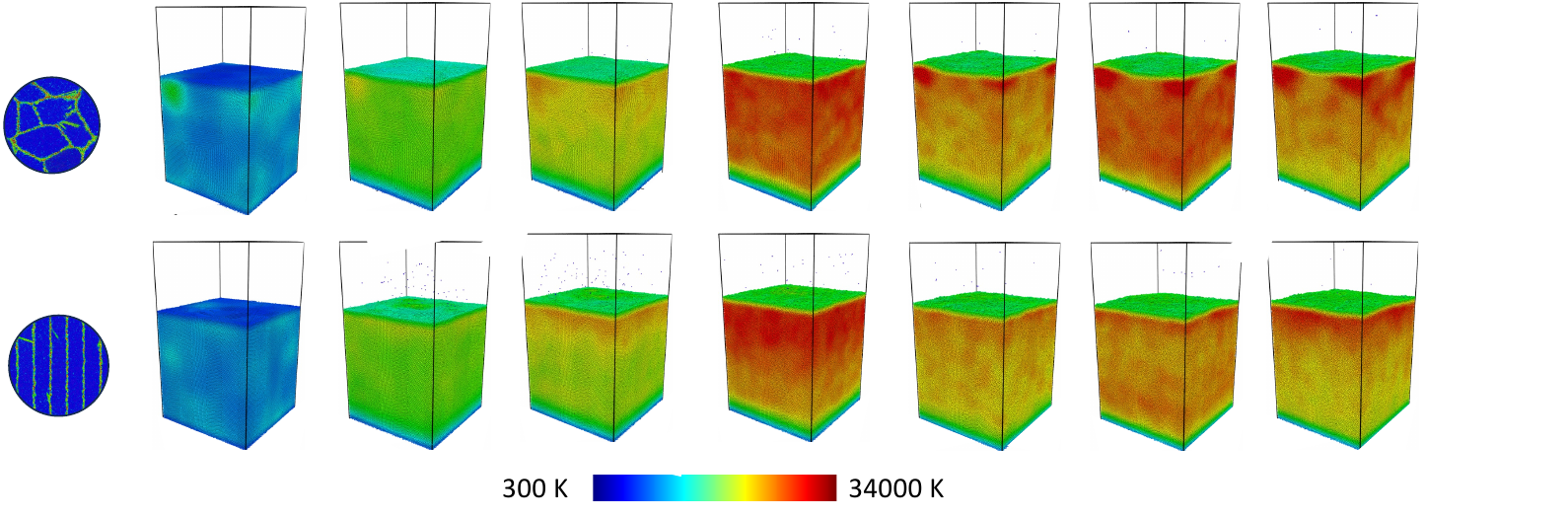}
\caption{Spatio-temporal evolution of temperature for different Au thin-film models irradiated with $ 1.28  J/cm^{2}$.}\label{Fig:SpatioTemporal_TemperatureEvolution}
\end{figure*}  

\textbf{Temperature Effects:}
Fig. \ref{Fig:SpatioTemporal_TemperatureEvolution} shows the spatiotemporal evolution of \textit{spatially averaged} atomic temperature for 1.28\thinspace$\si{\joule/\centi\meter}^2$.

Both SC thin films exhibited a nearly homogeneous temperature profile until 15\thinspace\si{\pico\second}, followed by a smooth temperature gradient around 20\thinspace\si{\pico\second}. 
Then, we observed a peak temperature distribution near 30\thinspace\si{\pico\second}. Later, after 30\thinspace\si{\pico\second}, the SC thin films exhibited homogeneous high-temperature\footnote{The time window for attaining peak temperature corresponds to Au's electron-phonon equilibration time (i.e., 20-30\thinspace\si{\pico\second}).} distribution.
In contrast, after 30\thinspace\si{\pico\second}, poly-NC thin films exhibited inhomogeneous temperature distribution with extremely hot (near the surface) and moderately hot zones (middle and lower region). Such relatively low-temperature zones correspond to expanding thin films during tension, evidenced by the spatiotemporal pressure distributions (refer Fig. \ref{Fig:SpatioTemporal_PressureEvolution}).

In this study, the poly-NC thin films recorded a peak globally averaged temperature (2300\thinspace\si{\kelvin}) relative to SC counterparts (1800\thinspace\si{\kelvin}) for higher applied laser fluences. 
In contrast, they exhibited a comparable global temperature trend at moderate and lower applied fluences. 
However, the difference in spatiotemporal temperature distribution (refer Fig. \ref{Fig:SpatioTemporal_TemperatureEvolution}) for SC and poly-NC thin films might stem from the following factors. In the case of SC, high energy deposited by an external laser pulse was utilized to effectively distort the crystal structure (i.e., in overcoming the cohesive energies of the solid system). In contrast, the poly-NC systems contained a significant volume fraction of grain boundary atoms (i.e., mostly incoherent atoms). Such low local atomic densities near GB and planar defects could facilitate atomic mobilization/displacements given sufficient energy, contributing to effective heating, thus reaching high global temperatures and substantial melting of thin films.
Furthermore, these observations become critical in predicting the lateral direction effects of laser irradiation, where the interaction of moderate regional fluence with microstructure features and preexisting defects plays a decisive role in determining the vertical and horizontal precision of nano machining.

Our results revealed that both homogeneous melting (nucleation of multiple liquid regions, which eventually grow) and heterogeneous melting (nucleation/expansion of liquid regions from free surface/grain boundary) were not only necessarily decided by the applied/deposited laser energy intensity but also by the underlying microstructure and crystallography.
To this end, the two crucial factors (i.e., energy and structure as discussed in section \ref{sec:latticeDistortion}) and their interplay determined the resulting thermodynamic indicators temperature and pressure, influencing the available driving forces.
Therefore, we would argue that interpreting the results solely based on thermodynamic factors without including the structural factors might not give the complete picture of ultrafast laser-metal interactions.  

\begin{figure*}
    \centering
    \begin{subfigure}[b]{0.5\textwidth}
      \includegraphics[scale=.395,trim={1.3cm 4.4cm 1cm 2.8cm},clip=true]{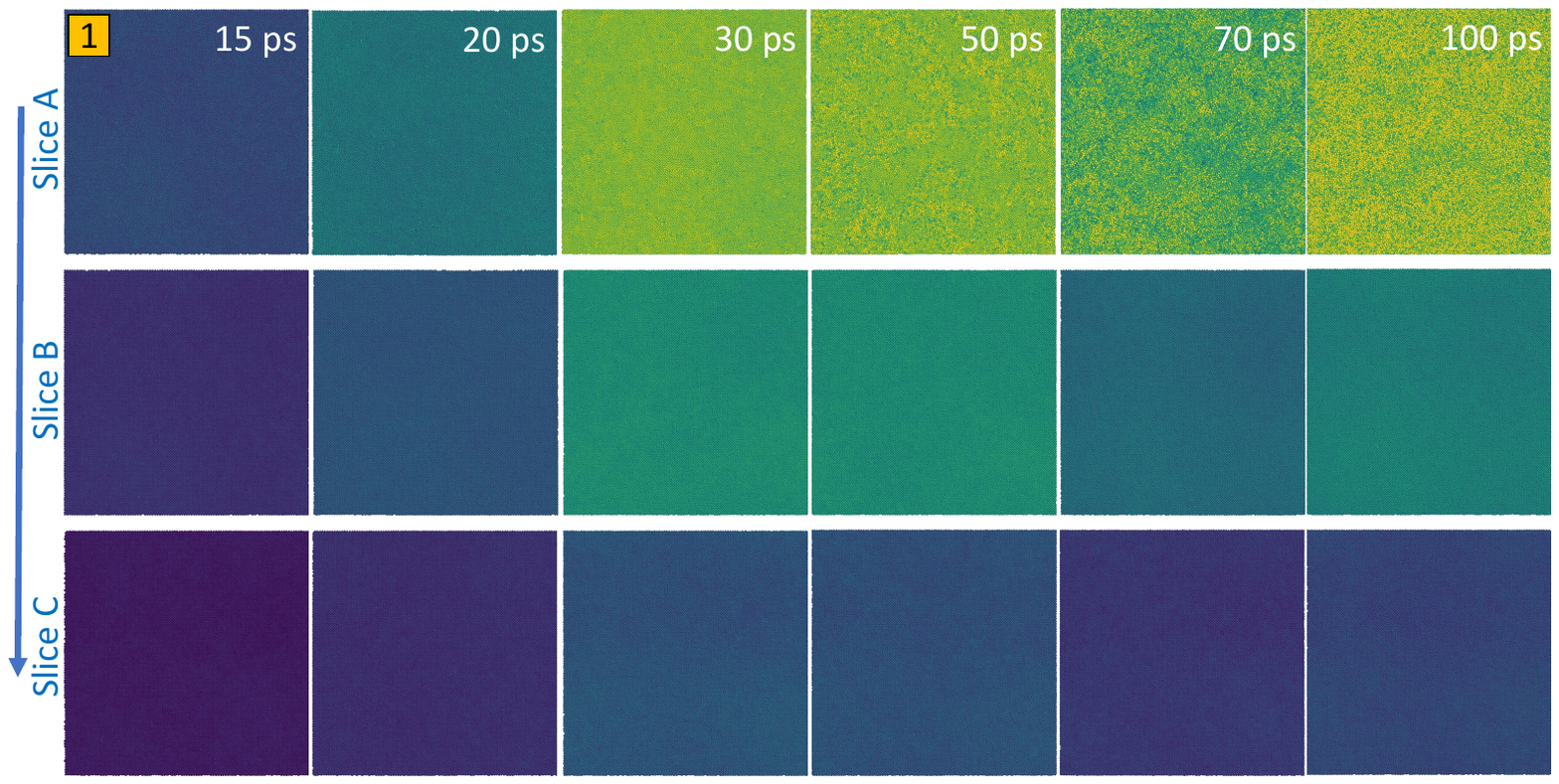}
    \end{subfigure}
    \begin{subfigure}[b]{0.5\textwidth}
      \includegraphics[scale=.395,trim={1.3cm 4.4cm 1cm 2.8cm},clip=true]{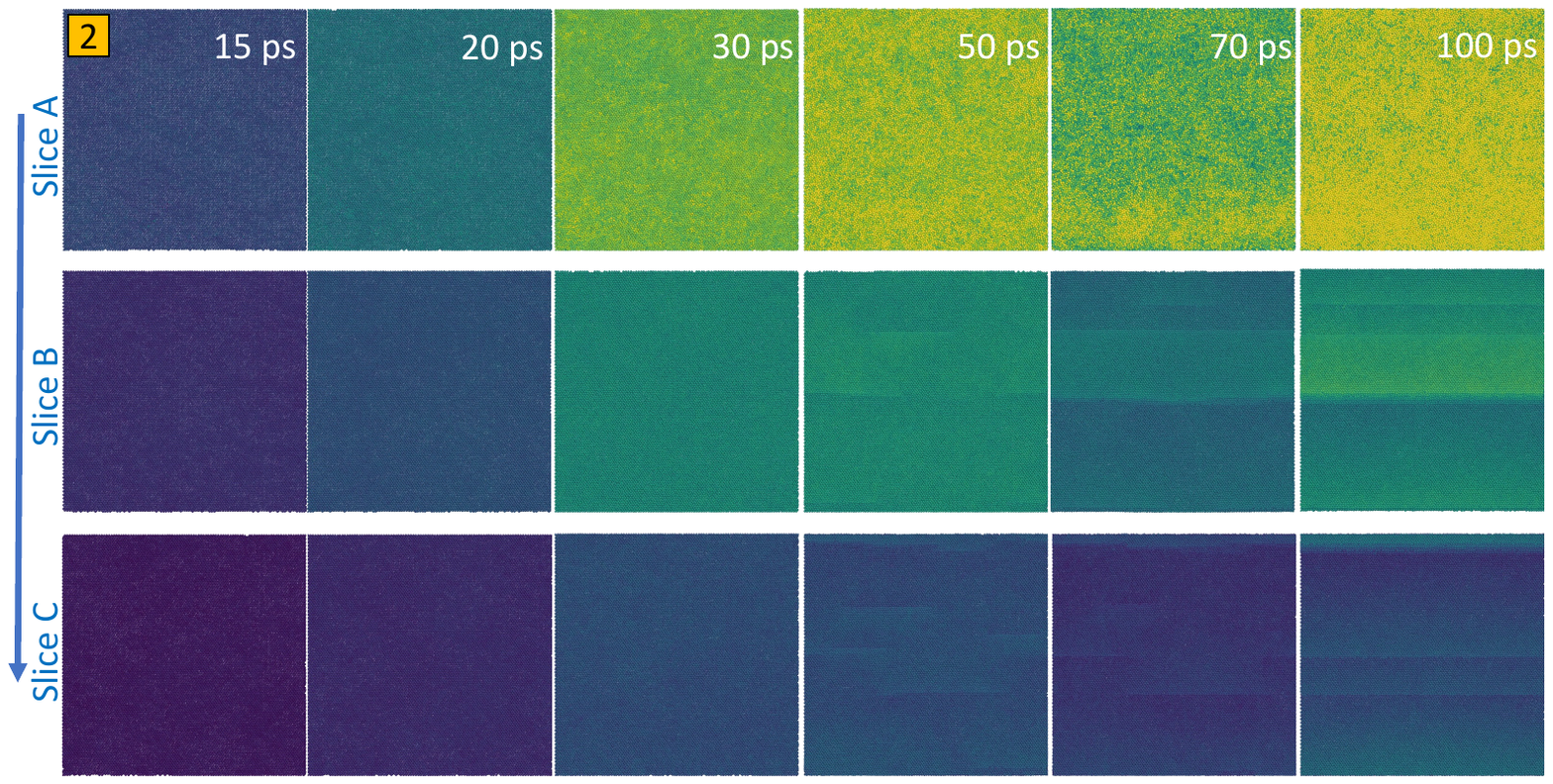}
    \end{subfigure}
    \begin{subfigure}[b]{0.5\textwidth}
       \includegraphics[scale=.395,trim={1.3cm 4.4cm 1cm 2.8cm},clip=true]{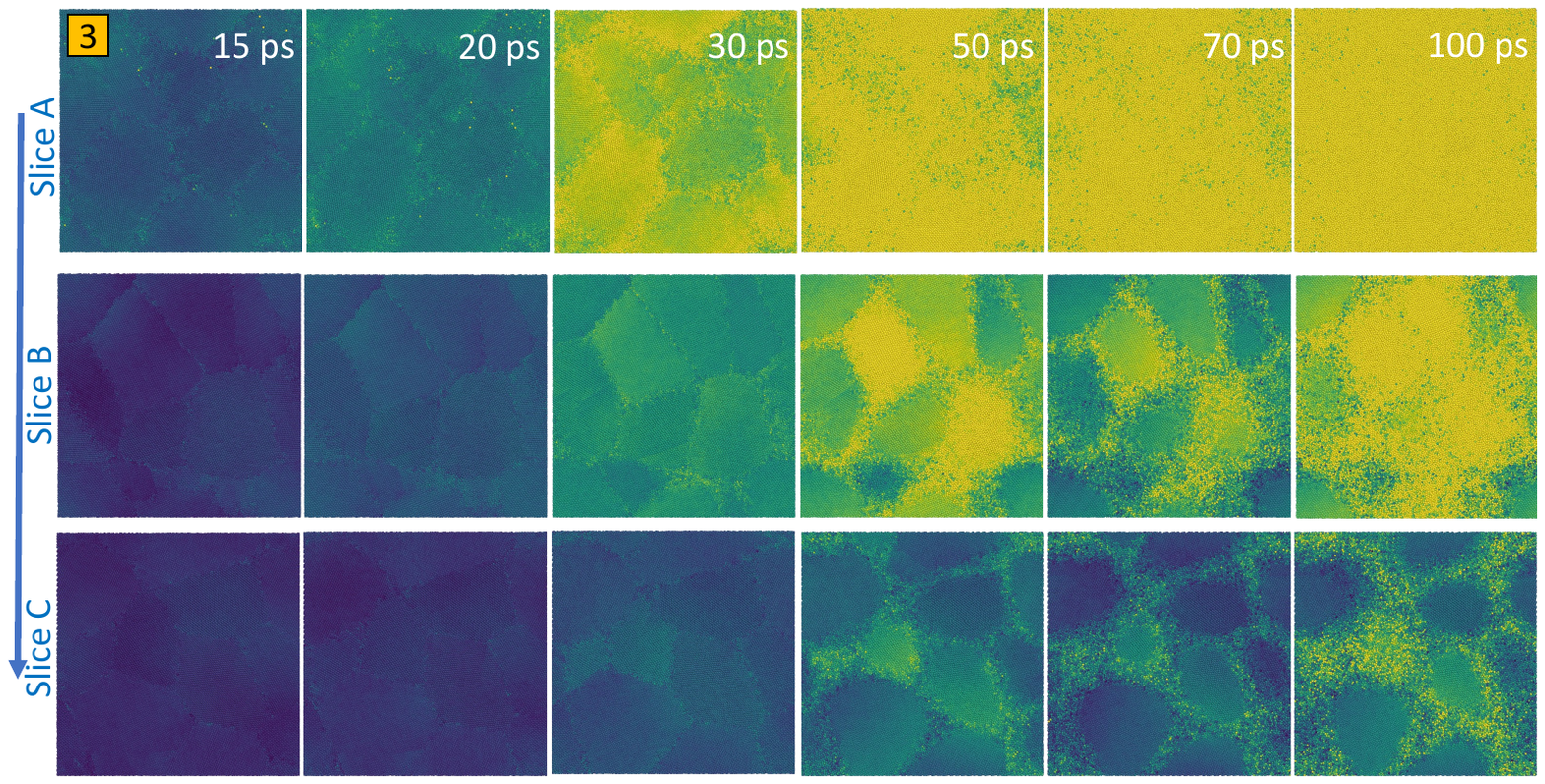}
    \end{subfigure}
    \begin{subfigure}[b]{0.5\textwidth}
       \includegraphics[scale=.395,trim={1.3cm 4.4cm 1cm 2.8cm},clip=true]{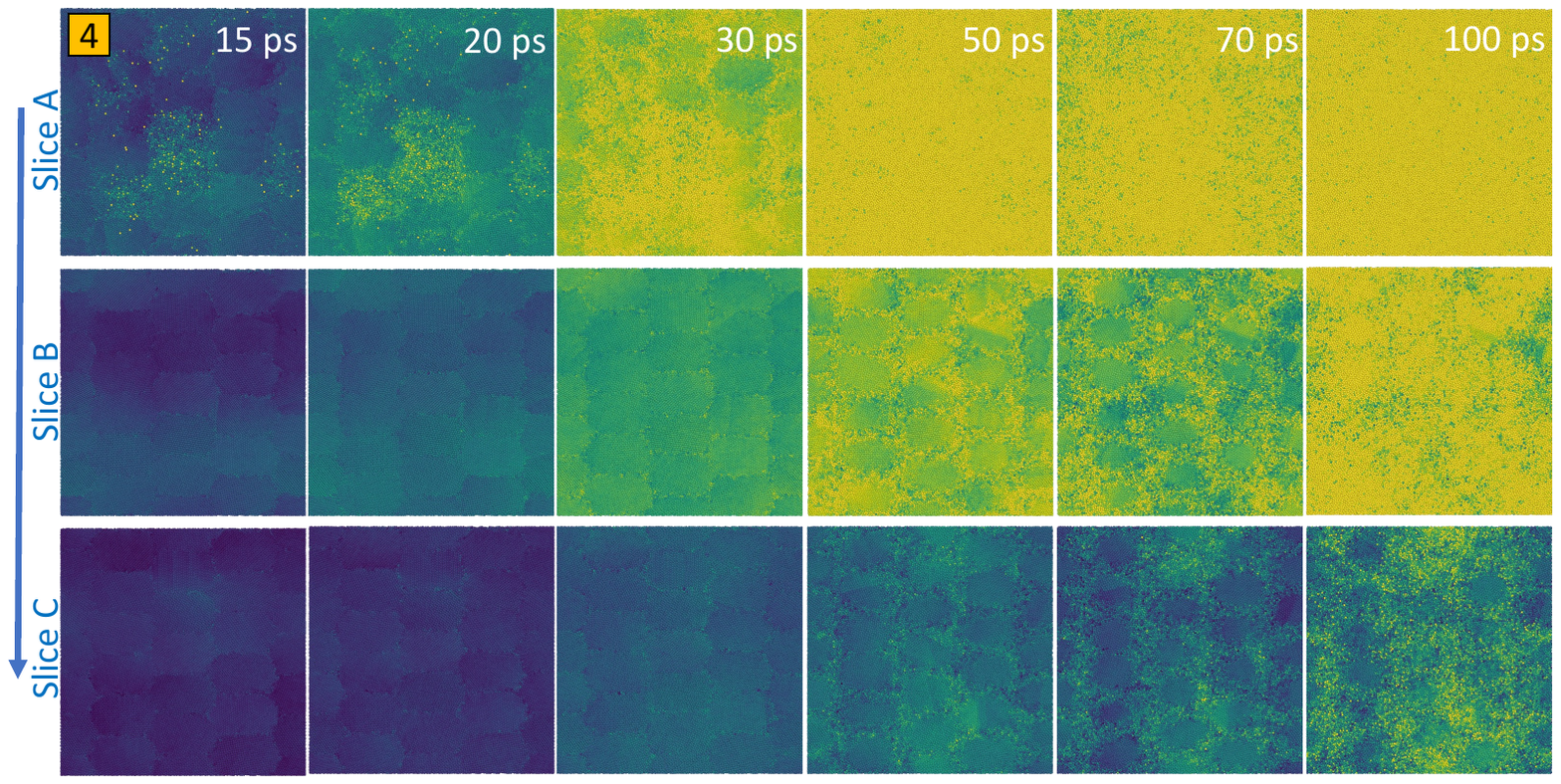}
    \end{subfigure}
    \begin{subfigure}[b]{0.5\textwidth}
        \includegraphics[scale=.395,trim={1.9cm 7.8cm 1.5cm 8.8cm},clip=true]{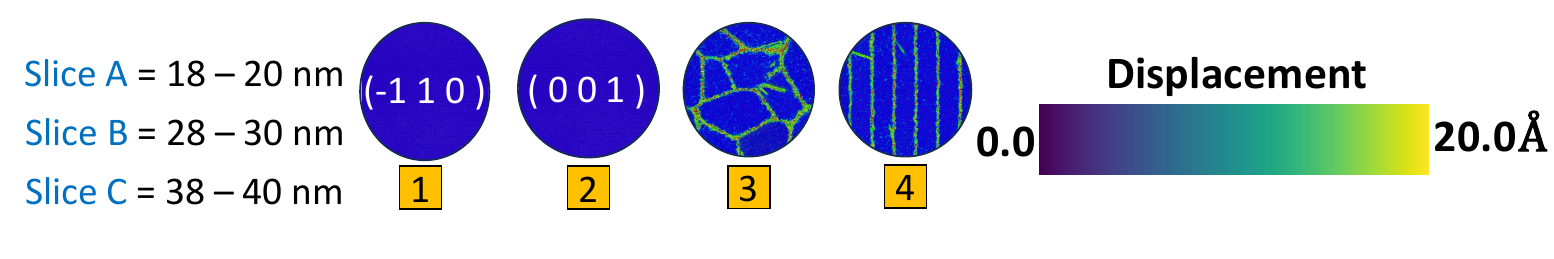}
    \end{subfigure}    
    \caption{Comparison of computed atomic displacements distribution at three different sliced regions (Slices: A, B, C) for different Au thin film models (Top to bottom): (1) SC\thinspace\hkl(-1 1 0), (2) SC\thinspace\hkl(0 0 1), (3) poly-NC (Random), (4) poly-NC (Columnar).}\label{Fig:AtomicDisplacementComparison}
\end{figure*}

\subsubsection{Laser-induced atomic displacements in thin films:}

Besides comparable thin film geometry and applied laser fluences, we observed a strong influence of underlying crystallographic orientation and microstructure topology. We elucidated these influences based on structural (i.e., atomic distortion, evolution of defects, and phase volume fraction) and quantitative global thermodynamic indicators (pressure and temperature). 
However, the atomic scale mechanisms behind the observed deviation in spatiotemporal quantities remain unexplored.
MD simulations provide access to such crucial information.
To this end, we computed the atomic displacements for three different sliced regions (each 2\thinspace\si{\nano\meter} thick) along the laser penetration direction at 18\thinspace\si{\nano\meter}, 28\thinspace\si{\nano\meter}, and 38\thinspace\si{\nano\meter}, respectively (i.e., along the film thickness X as shown in Fig. \ref{Fig:AtomisticModel}) for 1.28\thinspace$\si{\joule/\centi\meter}^2$.
These displacement values were computed w.r.t initial configuration (t = 0\thinspace\si{\pico\second}) before laser irradiation.

Fig. \ref{Fig:AtomicDisplacementComparison} shows the temporal evolution of atomic displacements for SC and poly-NC thin film models.
SC\thinspace\hkl(-1 1 0) thin film showed smoothly increasing atomic displacements over time, such that slice A (top) recorded some moderate displacement distribution after 30\thinspace\si{\pico\second}. In contrast, slices B and C showed nearly homogeneous displacement distributions with low magnitude. 
These observations validate that the lattice distortions were confined near the irradiation surface.
On the other hand, SC\thinspace\hkl(0 0 1) showed relatively more atoms exhibiting high displacements on slices A and B (owing to nucleated stacking faults as shown in Fig. \ref{Fig:StackingFault}), whereas slice C showed a similar displacement profile with lower magnitude.

In contrast to SC thin films, both Poly-NC thin films (i.e., random and columnar) displaced relatively high even at an early stage (e.g., at 15\thinspace\si{\pico\second}) and continued further after 30\thinspace\si{\pico\second}. 
By construction, the poly-NC (random) configuration's grain topology varies as we traverse along the depth. This explains the changing grain morphology at a given instance from top to bottom. In contrast, poly-NC (columnar) showed nearly uniform topology along the depth.
At the start, we recognize such underlying GB topology for slice A complying with the displacement profile (refer to slice A on rows 3 and 4 in Fig. \ref{Fig:AtomicDisplacementComparison}). However, the underlying GB topology pattern slowly disappears over time and along the depth (i.e., slices B and C), indicating the molten/disordered phase of the thin film, i.e., laser-induced phase transformation. 
In particular, at 100\thinspace\si{\pico\second}, slice A resulted in complete melting or distortion regardless of underlying microstructure topology. 
   
To generalize, we observed two interesting aspects: 1.) the laser-induced atomic displacements/distortions penetrate deeper into the thin film, and 2.) some crystalline regions showed a high degree of atomic displacement over the others demarcated by grain boundaries.
The former aspect delineates that poly-NC thin films exhibit relatively high laser-induced deformation even at lower applied laser fluence, in contrast to SC thin films. The latter aspect sheds light on the critical role of local crystal orientations and grain topology in contributing to heterogeneous nucleation/melting. 
Altogether, poly-NC (columnar) exhibited more displacement both in time and depth.

\subsection{General Discussions:}

In metals, ultrafast heating during femtosecond laser irradiation leaves the system in a superheated state such that solid-liquid transformation occurs well above the equilibrium melting temperature \cite{forsblom2005superheated}.
Recently, \cite{arefev2022kinetics} investigated the laser-induced melting of thin gold films using hybrid TTM-MD and reported homogeneous melting mechanisms for all the energies. Furthermore, they noted that heterogeneous melting cannot be observed using the variation of electron-phonon coupling strength, particularly near the melting threshold.
However, our simulations revealed both homogeneous and heterogeneous melting mechanisms for higher and lower applied fluences studied herein. Our simulation results agree with the experimental work \cite{mo2018heterogeneous}, which reported a similar transition in melting mechanisms for gold thin films.
We attribute our observations of such transition to two factors: 1.) incorporation of microstructure features in thin films and 2.) electron temperature-dependent interatomic potential. 
In the past, \cite{lin2010molecular} also attempted to study nanocrystalline thin films but did not systematically account for topology and crystallographic orientations. Even in that case, every observation was made in the light of thermodynamics but not from the associated nanomechanics, despite the other critical differences with our work, namely thin film dimensions, interatomic potential (underprediction of melting temp), applied laser profile, pulse duration, thermophysical parameters, and boundary conditions.

During ultrafast irradiation of metallic thin films, the deposited high energy introduces a strong temperature gradient, unfolding a sequence of events as follows: 1. heating of electrons and phonons, 2. heat transfer from electronic to lattice subsystem, 3. laser-induced thermomechanical pressure, 4. volume expansion, 5. phase transformation, 6. laser-induced deformation/ablation.
Consequently, investigating the material under such an extreme environment remains crucial for delineating the origins of several non-equilibrium events, e.g., rapid heating, melting, and laser-induced microstructure changes.
Typically, the investigation of ultrafast laser-metal interactions focuses mainly on applied laser process parameters and thermophysical properties.
In addition to these physical conditions, we emphasized that the underlying thin film microstructure topology and crystallographic orientations could also play a decisive role.
We based our emphasis on the observation that even under the same laser processing parameters, thermophysical conditions, and identical target thin-film dimensions, the resulting laser-induced deformation behavior varies. 

Our simulations revealed that initial grain topology and orientations influenced the laser-induced stresses, thus affecting the thin film expansion behavior. 
We delineated the crucial nanomechanical aspects of laser-induced deformation, characterized by varying defect evolutions (activated events) and the extent of lattice distortions (melting mechanisms).
Consequently, even under identical irradiation conditions, some thin films' microstructure models showed higher melting/lattice distortion than others. 
Our simulation results validated that the activation of melting mechanisms (i.e., homogeneous/heterogeneous) depends not only on the energetic factor (i.e., applied laser fluence) but also on the underlying microstructure factor (i.e., crystallographic orientations, grain boundaries, and microstructure defects) and their interplay.

Investigating ultrafast laser-metal interactions using the TTM-MD simulation approach offers an atomic-level understanding of the role of underlying thin-film crystallographic orientations, microstructure, and topology. 
To this end, the spatiotemporal evolution of defects, crystalline phases, and atomic displacements elucidates the atomic-scale mechanisms behind the observed melting and high proportion of lattice distortion. 
In contrast, thermodynamic quantities such as pressure and temperature elucidate only the driving forces behind the activated mechanisms and their interplay.

Among the four thin film models investigated, poly-NC (columnar) showed extensive melting characterized by loss of crystallinity (i.e., steady decline in the fcc phase).
Consequently, our results revealed that the two crucial factors (i.e., energy and structure) and their interplay determine the spatiotemporal temperature and pressure distributions, thus, the resulting melting mechanisms.
The laser-induced atomic displacement analysis revealed that poly-NCs undergo deeper lattice distortions than SC thin films, even at lower applied laser fluence. 
Some crystalline regions showed higher atomic displacement than others demarcated by grain boundaries, emphasizing the critical role of local crystal orientations and grain topology in contributing to heterogeneous nucleation/melting.
The opportunities provided by ultrafast laser-metal interactions are twofold: 1.) selective/targeted material microstructure modification, 2.) enabling the investigation of structure-property relationship under extreme thermomechanical conditions. 
Regardless of identical thermophysical, laser processing conditions and thin-film geometry, our simulations revealed that the extent of lattice heating and ease of deformation depend critically on the thin film’s local crystallographic orientations and microstructure topology.

\section{Conclusions}

Progress in the emerging fields of atomic and close-to-atomic scale manufacturing is underpinned by enhanced precision and optimization of laser-controlled nanostructuring/nano machining, which requires a profound understanding of ultrafast laser-thin film interaction at atomic resolution.
In particular,  thin films’ crystallographic orientations and microstructure effects become critical for optimizing the laser-metallic thin film interactions.
Using hybrid TTM-MD, we simulated ultrafast laser-metal interactions under different applied laser fluences (low and high-fluence regimes) for four thin film models (i.e., SC\thinspace\hkl(-1 1 0), SC\thinspace\hkl(0 0 1), poly-NC (random), poly-NC (columnar)) with varying microstructures and crystallographic orientations.

For higher applied laser fluences, both SC and poly-NC thin films underwent homogeneous melting characterized by rapid nucleation of several liquid-like highly disordered atomic regions. Then, for the lower applied fluences, we observed heterogeneous melting as the liquid/disorder region confined near the surface for SC thin films and extensive grain boundaries mediated melting for poly-NC thin films.
Herein, for higher applied laser fluences, thin film expansions nucleated multiple nanovoids that grew and coalesced to form cavitation bubbles.
These cavitations acted as pressure-relieving mechanisms and were sensitive to underlying initial thin-film microstructure and crystallographic orientations.
SC thin films withstood relatively higher laser-induced pressure as they utilized it for defect nucleation than poly-NC thin films with preexisting grain boundary dislocations. 
The onset of dislocations and planar defects as pressure-relieving mechanisms delineated the utilization of deposited thermal energies, which is beneficial in delaying the fragmentation (e.g., void creation) of thin films.

Microstructure features, namely grain size, grain topology, and local crystallographic orientation, controlled the rate and extent of lattice disorder evolution and phase transformation, particularly at lower applied fluences. 
For a comprehensive investigation in the future, we also deem it is important to consider other important aspects, namely simulating long-time  microstructure effects, experimentally informed laser source profile, geometric dimension of thin films (lateral directions), inclusion of substrate effects, and thin film thickness variations.
Combining nanomechanical and thermodynamic aspects will enhance the understanding of such complex laser-metal interaction at atomic resolution. 
In short, these two factors (i.e., energy and structure) critically determine the degree of lattice distortion and, thus, the resulting ablation profile.
For the first time, our simulations provided comprehensive insights encompassing the nanomechanical and thermodynamic aspects of ultrafast laser-metal interactions and their interplay at atomic resolution. 
Microstructure-aware/informed thin film fabrication and targeted defect engineering could improve the precision of nanoscale laser processing and potentially emerge as an energy-efficient optimization strategy.

\section{Declaration of Competing Interest}
The authors declare that they have no known competing financial interests or personal relationships that could have appeared to influence the work reported in this work.

\section{Acknowledgments}

The authors gratefully acknowledge funding from the German Research Foundation (DFG) - Grant No. 469106482 (SA2292). Furthermore, fruitful discussions with Michael Budnitzki, Markus Olbrich, Theo Pflug, and Sergei Starikov are gratefully acknowledged.

\section{Data availability}
The raw/processed data required to reproduce these findings cannot be shared at this time as the data also forms part of an ongoing study.

\printcredits

\end{document}